\documentclass[twocolumn,aps,pre]{revtex4-1}

\usepackage[pdftex]{graphicx}
\usepackage{dcolumn}
\usepackage{bm}
\usepackage{hyperref}

\usepackage{subfigure}
\usepackage{amsmath, amssymb}
\usepackage{enumitem}
\usepackage{color, colortbl}
\usepackage{afterpage}
\usepackage{url}
\usepackage{multirow}
\usepackage{rotating}
\usepackage{hyperref}
\usepackage{ulem}
\usepackage{braket}
\usepackage{xr}
\usepackage{soul}

\newcommand*{\card}[1]{|#1|}
\newcommand*{\subs}[1]{\mbox{\tiny $#1$}}

\begin{document}

\title{Disentangling co-occurrence patterns in n-partite ecosystems}

\author{Albert Sol\'e-Ribalta}
\affiliation{Internet Interdisciplinary Institute (IN3), Universitat Oberta de Catalunya, Barcelona, Catalonia, Spain}

\author{Claudio J.~Tessone}%
\affiliation{URPP Social Networks, Universit\"at Z\"urich, Switzerland}

\author{Carlo G.~Ferrari}
\affiliation{Departamento de Matem\'atica, Facultad de Ciencias Exactas y Naturales, Universidad de Buenos Aires, Argentina}
\affiliation{Instituto de Investigaciones Matem\'aticas Luis A. Santal\'o IMAS-UBA-CONICET, Buenos Aires, Argentina}

\author{Javier Borge-Holthoefer}
\affiliation{Internet Interdisciplinary Institute (IN3), Universitat Oberta de Catalunya, Barcelona, Catalonia, Spain}
\affiliation{Institute for Biocomputation and Physics of Complex Systems (BIFI), Universidad de Zaragoza, Zaragoza, Spain}

\date{\today}

\begin{abstract}
The need to harmonise apparently irreconcilable arrangements in an ecosystem --nestedness and segregation-- has triggered so far different strategies. Methodological refinements, or the inclusion of behavioural preferences to the network dynamics offer a limited approach to the problem, since one remains constrained within a 2-dimensional view of an ecosystem, i.e. a presence-absence matrix. Here we question this partial-view paradigm, and show that nestedness and segregation may coexist across a varied range of scenarios. To do so, we rely on an upscaled representation of an ecological community as an $n$-partite hypergraph, inspired by Hutchinson's high-dimensional niche concept and the latest trends on ecological multilayer networks. This yields an inclusive description of an ecological system, for which we develop a natural extension of the definition of nestedness to larger dimensional spaces, revealing how competitive exclusion may operate regardless of a highly nested bipartite system.

\end{abstract}

\maketitle



\section{\label{sec:level1}Introduction}

The characterisation of species co-occurrence patterns is a central question in ecology to understand not only which mechanisms enable the assembly of communities, but also how these will behave under changing environmental conditions. Community ecology largely agrees at large on the fact that such patterns are non-random. The debate remains, however, to sort out what non-random patterns should be observed.

In particular, two opposing architectures of species-species community organisation have been stubbornly observed, namely nestedness and segregated species co-occurrence. Nestedness is a very common property in empirical mutualistic\cite{bascompte2003nested} networks, but not only\cite{lewinsohn2006structure,burns2007network,valtonen2001structure}. In systems displaying nestedness, specialists (species with few partners) tend to interact with subsets of the mutualistic partners of generalists (species with many partners). The emergence of this frequent structural arrangement has ignited much research, in order to investigate the relation between the species network properties and its dynamical consequences, in terms of system stability and biodiversity \cite{bastolla2009architecture,suweis2013emergence}. Of course, the presence of nestedness implies a high niche overlap, clearly at odds with the Volterra-Gause competitive exclusion principle\cite{gause34,hardin60,macarthur1967limiting}, which states that two species with equivalent niches cannot exist locally and the one with lower fitness is convicted to extinction. Ultimately,  strong competition should result in checkerboard-like patterns\cite{diamond1975assembly}, produced by pairs of species with mutually exclusive ranges --a sort of anti-nestedness\cite{almeida2007nestedness}.

Both organisation principles have received an immense amount of empirical and theoretical support. In the middle ground, attempts to reconcile nestedness and negative species co-occurrence (segregation) have come either from a methodological stance (e.g. improving and standardising metrics and null modelling techniques\cite{ulrich2007disentangling}), or embedding behavioural preferences to the network dynamics, e.g. adaptive foraging\cite{valdovinos2013adaptive,suweis2013emergence,valdovinos2016niche} and/or spatio-temporal shifts to promote species coexistence\cite{connell80,kronfeld2003partitioning,castro2010assessment}. Little has been done, however, to embrace the problem respecting the dimensionality of the system under study (but see Ulrich {\it et al.}\cite{ulrich2017comprehensive}), which corresponds to Hutchinson's inclusive niche concept\cite{hutchinson1957}, formalised as a region in an $n$-dimensional hypercube, i.e. a multi-dimensional space where each dimension represents a resource. In particular, we note that, given the limitations to access concurrent geographical, temporal and species interaction data, almost all efforts have been placed on bipartite networks (species-species or species-site) which stand as {\it flat} (projected) views of ecological communities, where one or more dimensions have been collapsed.

To overcome these constraints, recent advances in the field suggest the benefits of considering higher-dimensional connectivity\cite{sander2015can,kefi2016structured,pilosof2017multilayer,garcia2018multiple}, clearing the way for a finer approach to structure-dynamics interplay\cite{gracia2017joint} in ecological networks. Along this line, we show here that high nestedness and low species co-occurrence are compatible within a larger dimensional space, and their concurrency can be parsimoniously explained when studied under the lens of higher-dimensional networks and a derived new structural pattern: $n$-dimensional nestedness.

\section{Results}
\label{sec:res}

\paragraph{Introducing $n$-dimensional nestedness.}
Nestedness is a classical structural measure used to study the species composition among isolated habitats (species-site), or to describe two disjoint sets of interacting species\cite{ulrich2009consumer}, e.g. plants and animals. In either case, a bipartite network is the convenient mathematical object to consider. However, along the lines of Hutchinson\cite{hutchinson1957}, the bipartite approach provides only a partial view --a projection of the higher dimensional space-- of the full ecological system whenever the dimensionality of the ecological system is larger than two\cite{eklof13}. To respect the dimensionality of the system, we require a similar high-dimensional structure, namely an $n$-partite hypergraph, where the set of all dimensions (species, habitats, time, etc.\cite{owen2015spatially}) can be modelled. 

\begin{figure*}[t]
    	\begin{tabular}{llll}
	        \multicolumn{4}{l}{{\bf (A)} }\\		
                \includegraphics[width=0.23\textwidth]{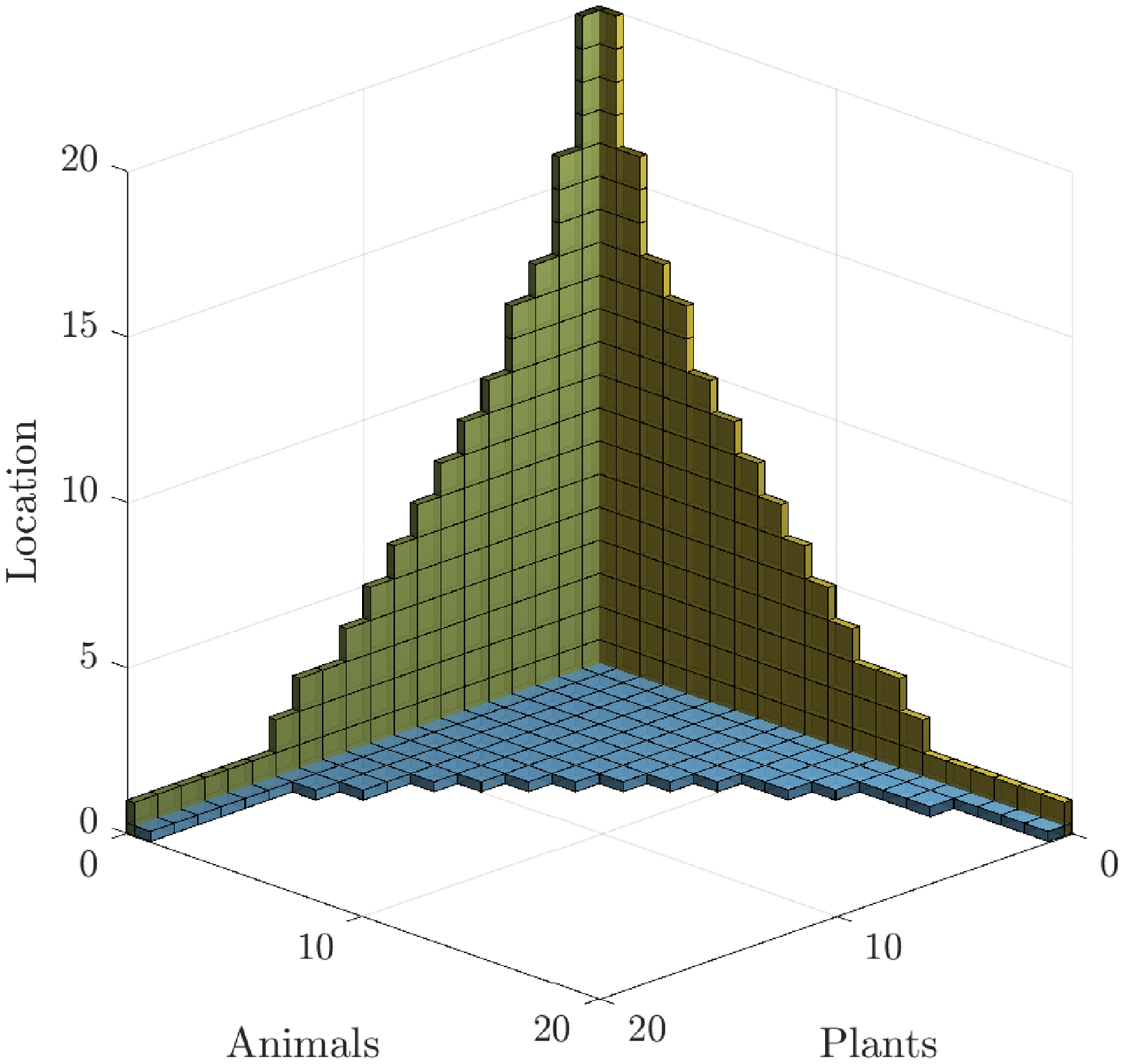} 
                & \includegraphics[width=0.23\textwidth]{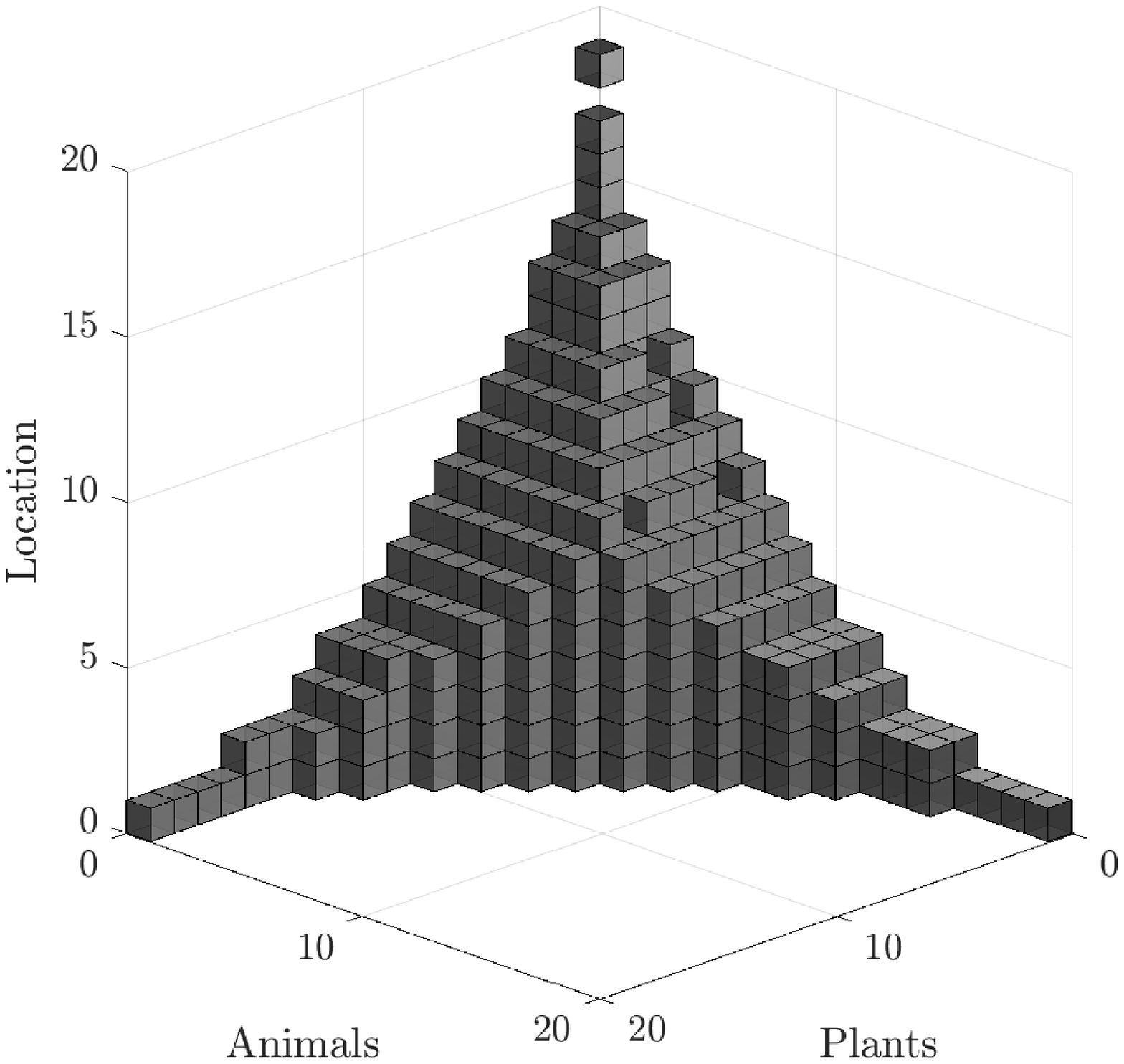} 
                & \includegraphics[width=0.23\textwidth]{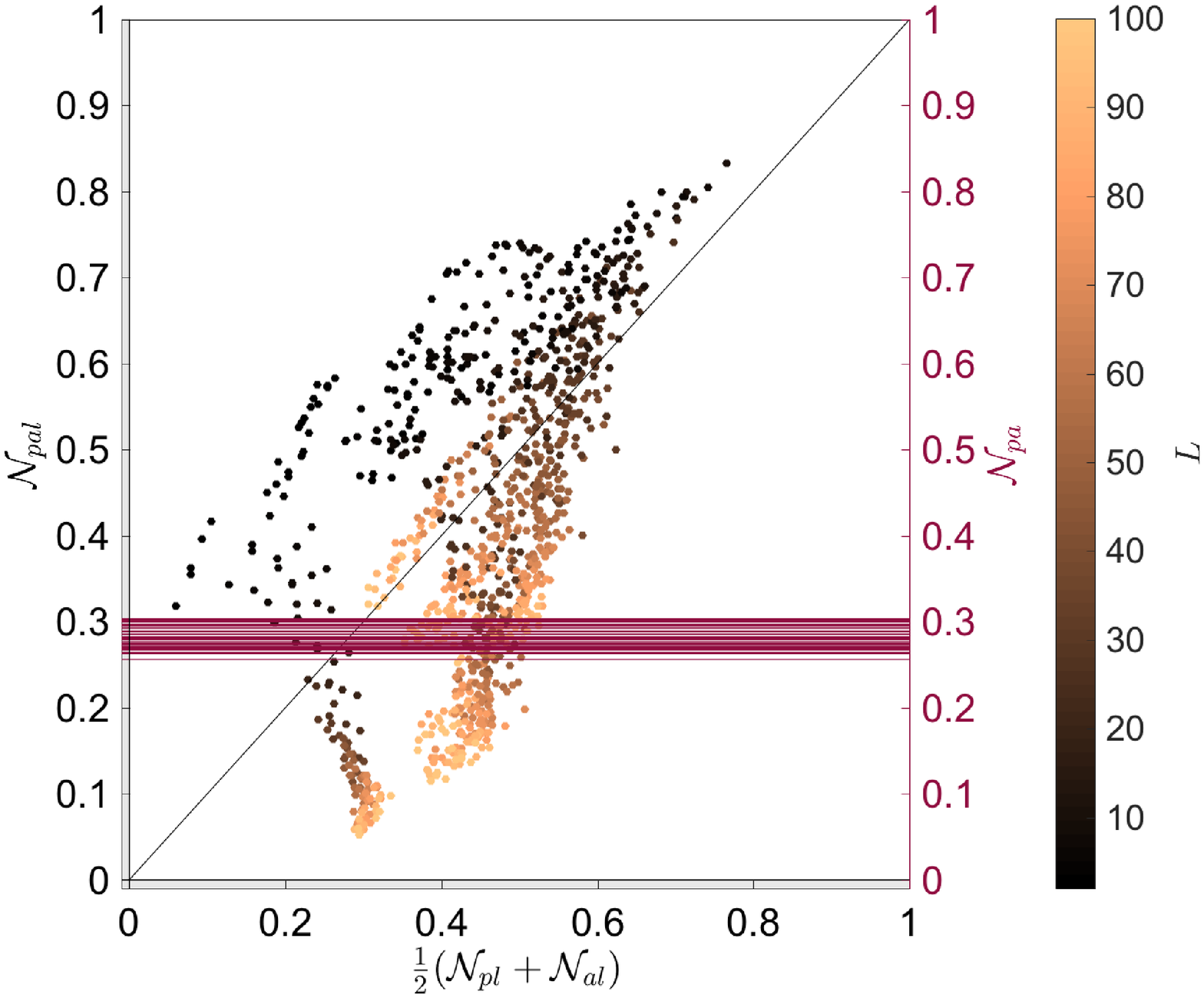}    
                & \includegraphics[width=0.21\textwidth]{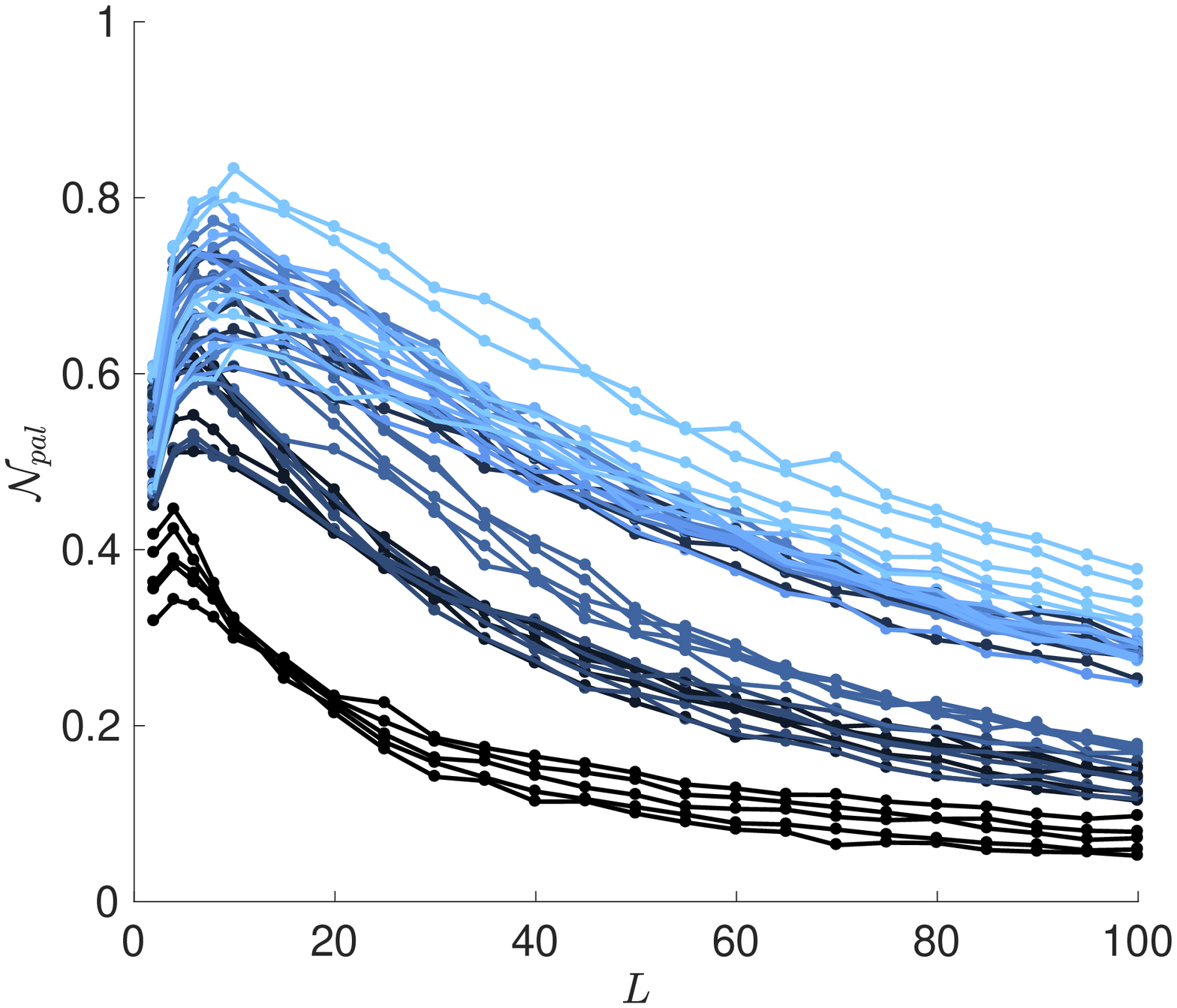}    \\
	        \multicolumn{4}{l}{{\bf (B)} }\\		
                \includegraphics[width=0.23\textwidth]{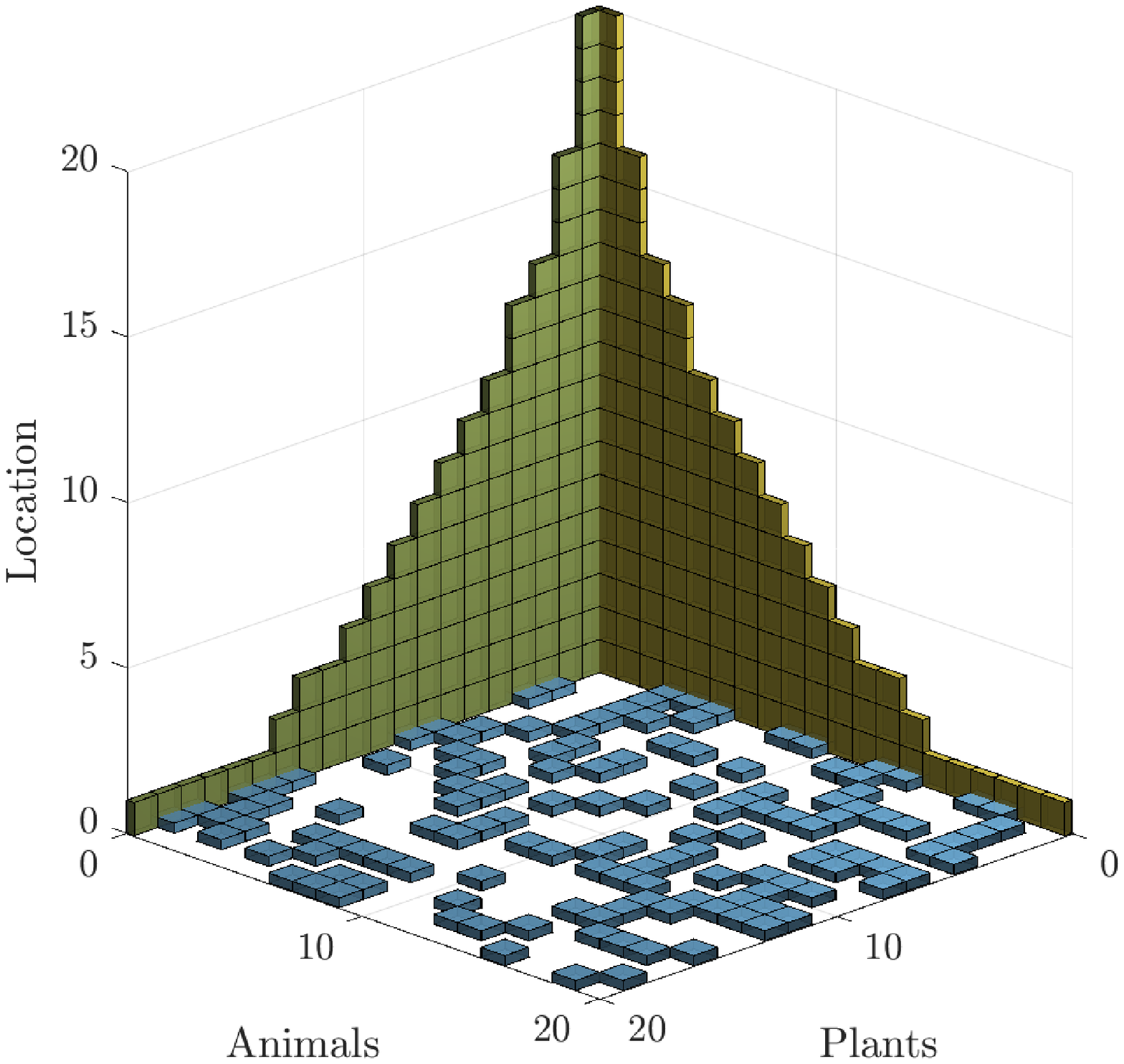} 
                & \includegraphics[width=0.23\textwidth]{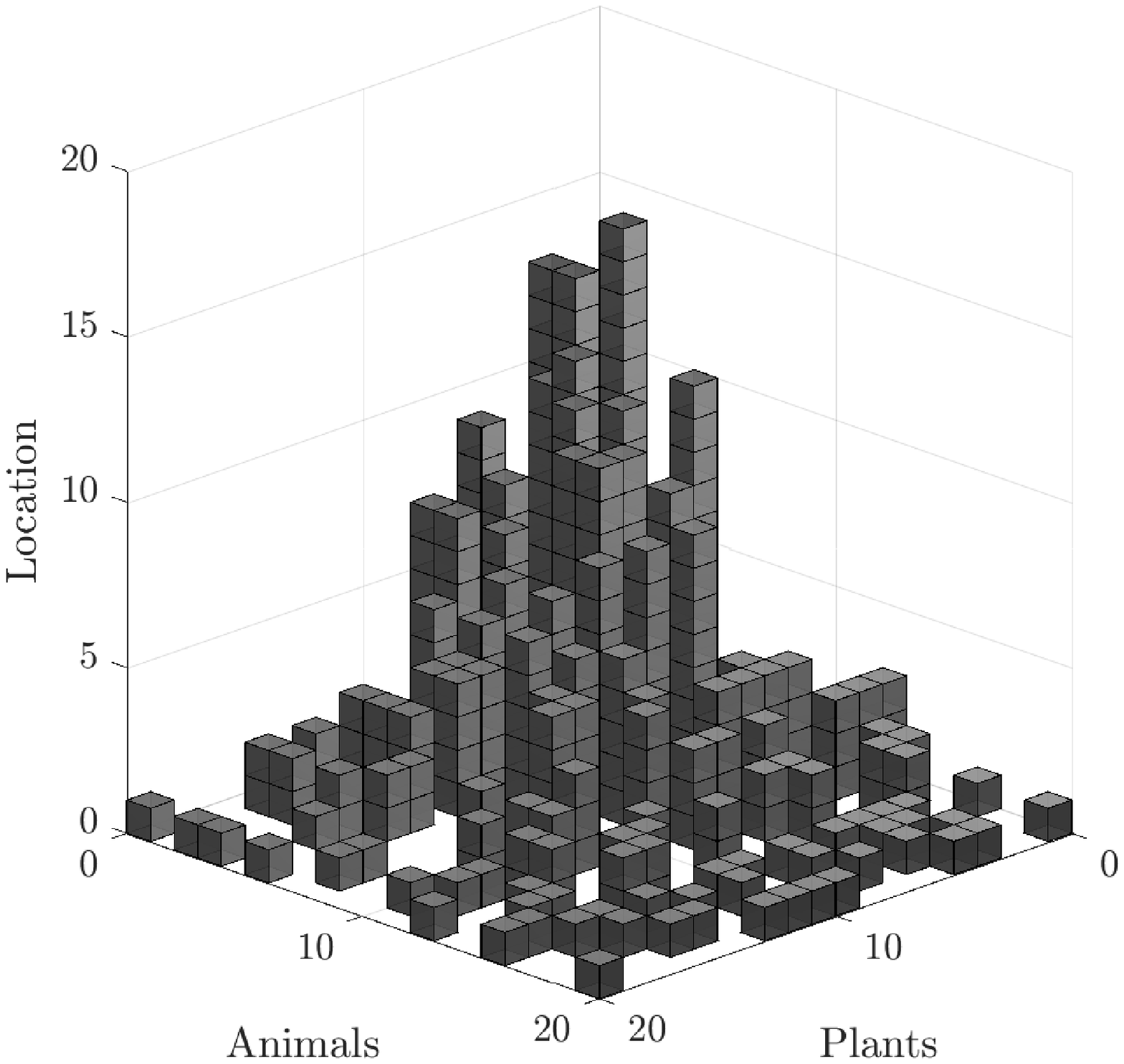} 
                & \includegraphics[width=0.23\textwidth]{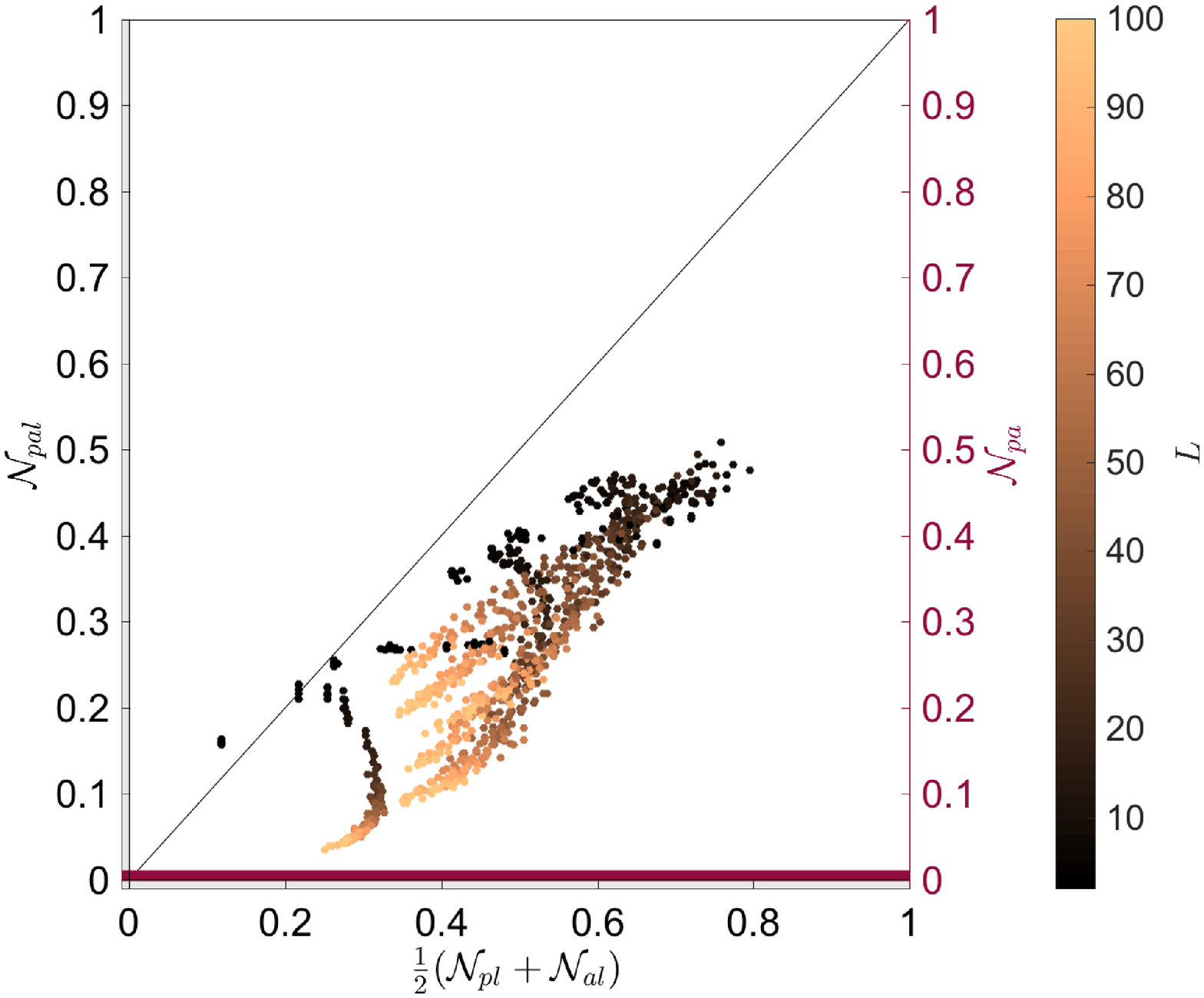}    
                & \includegraphics[width=0.21\textwidth]{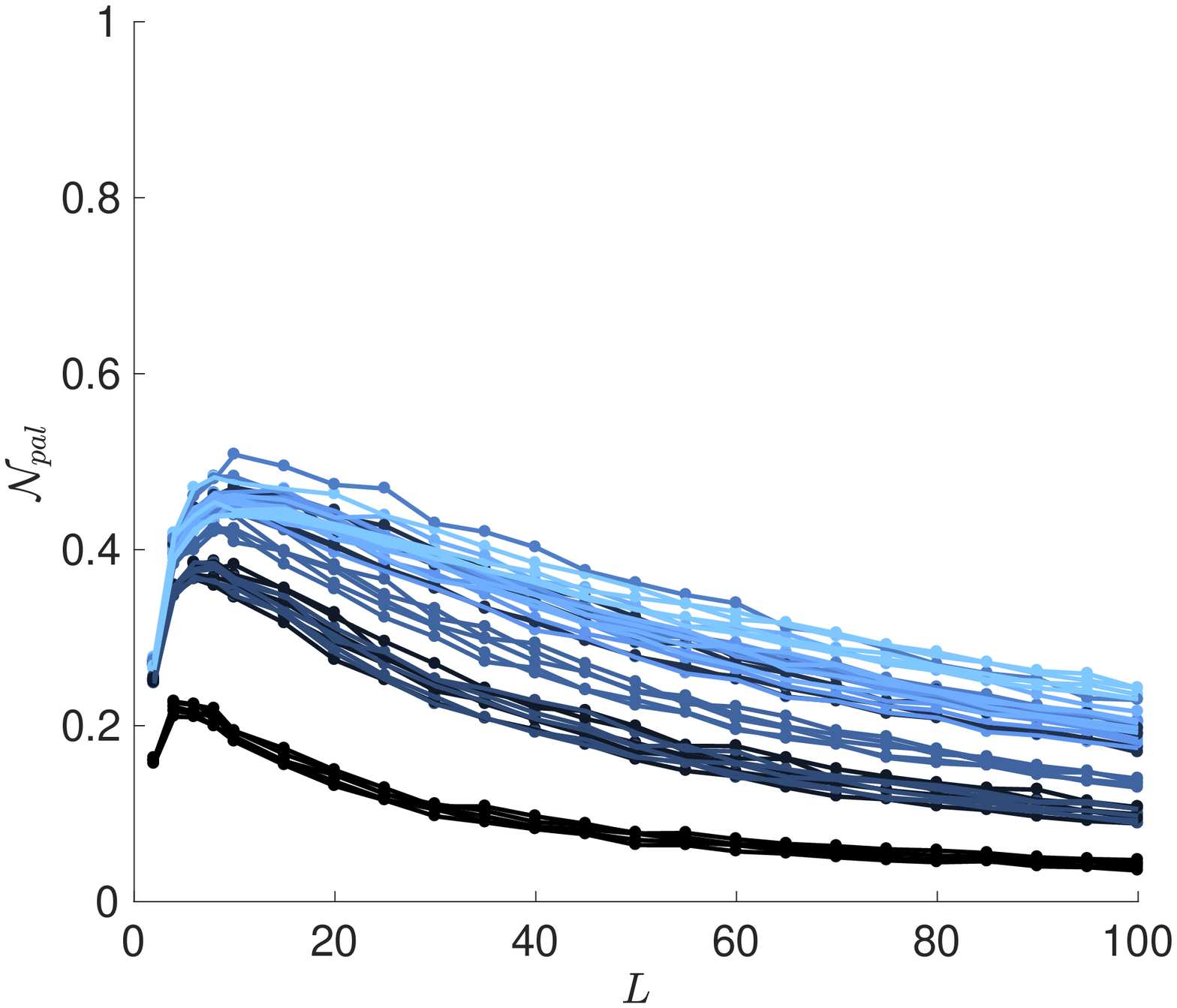}    \\
	        \multicolumn{4}{l}{{\bf (C)} }\\		
                \includegraphics[width=0.23\textwidth]{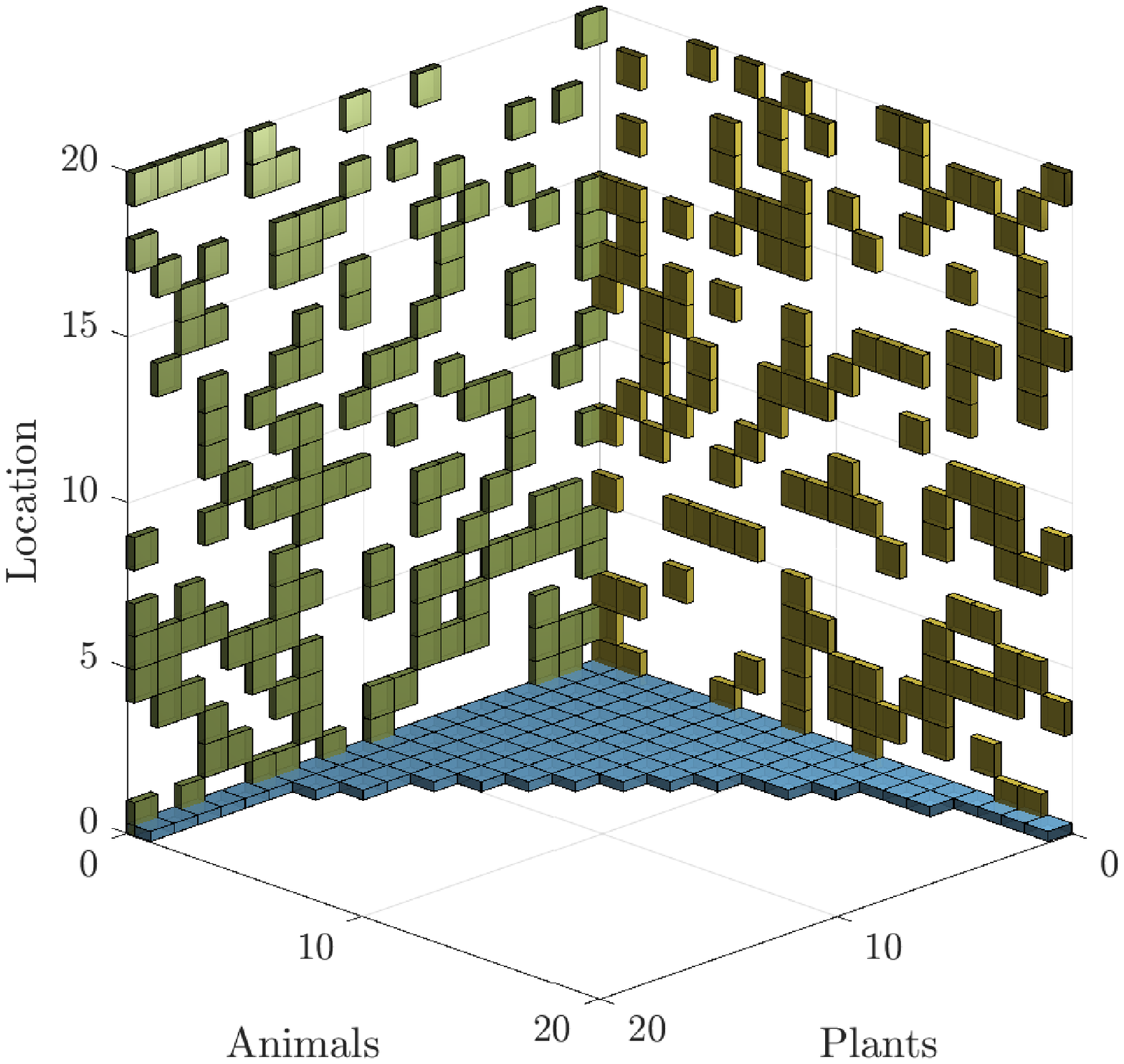} 
                & \includegraphics[width=0.23\textwidth]{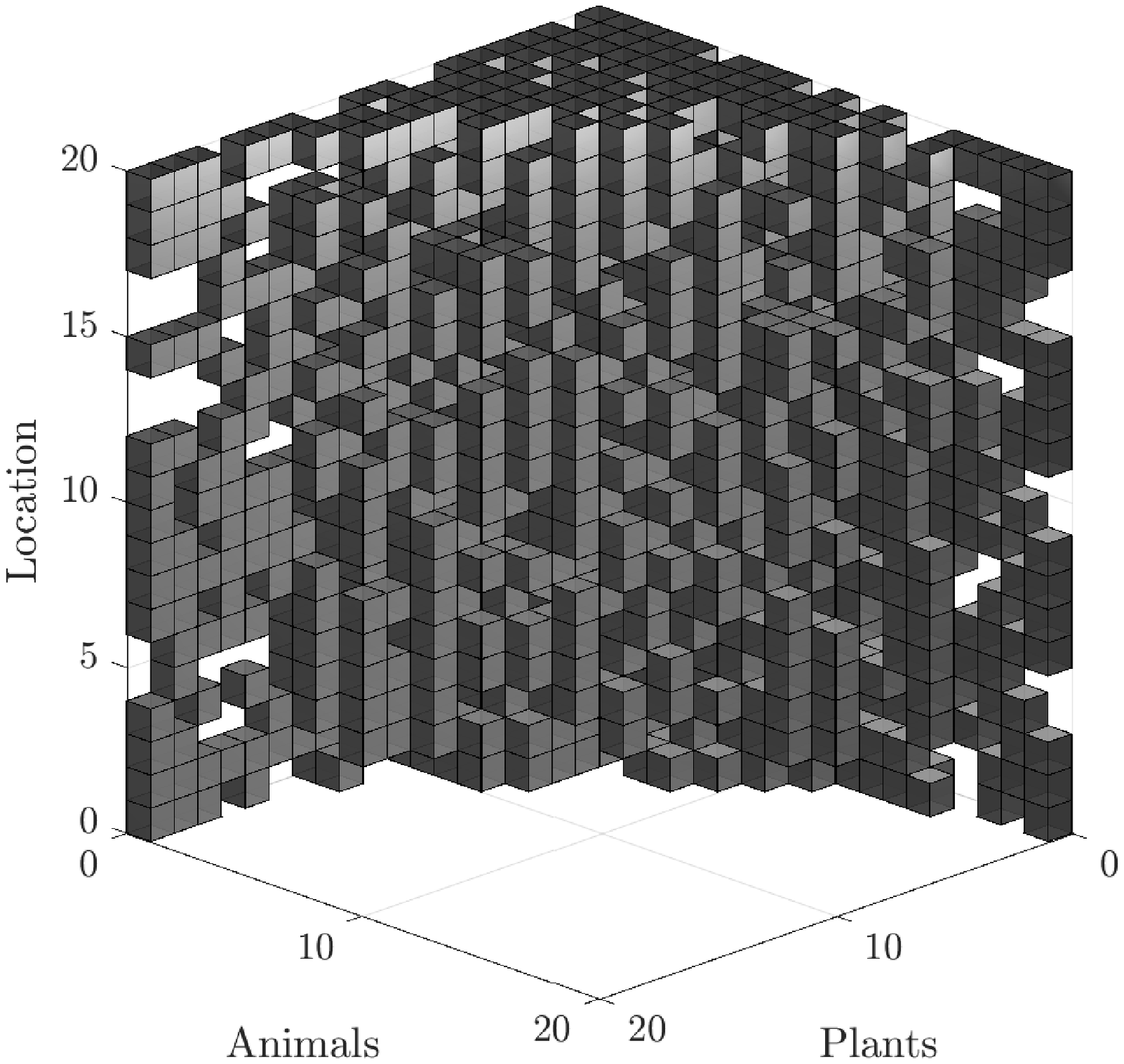} 
                & \includegraphics[width=0.23\textwidth]{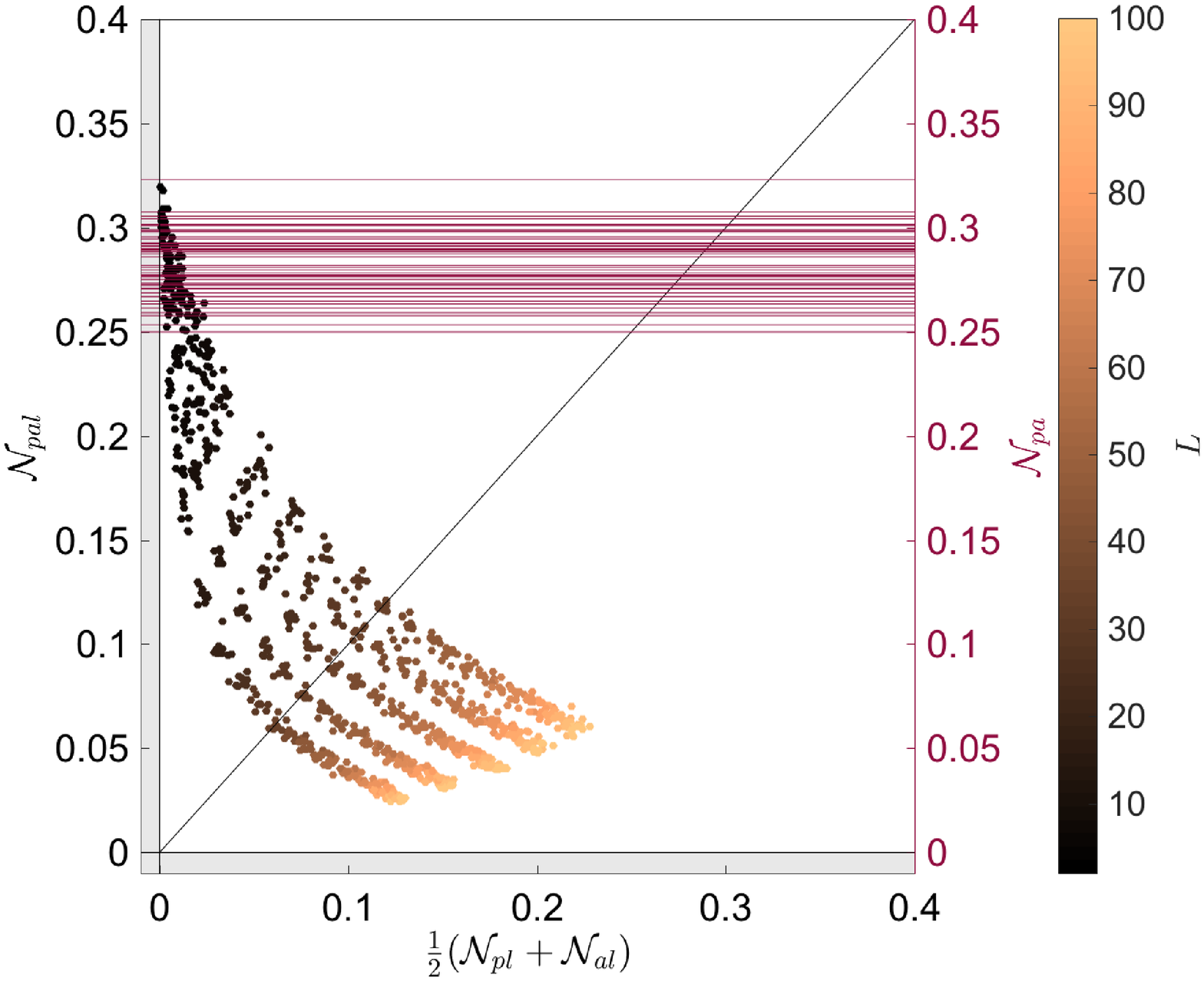}    
                & \includegraphics[width=0.21\textwidth]{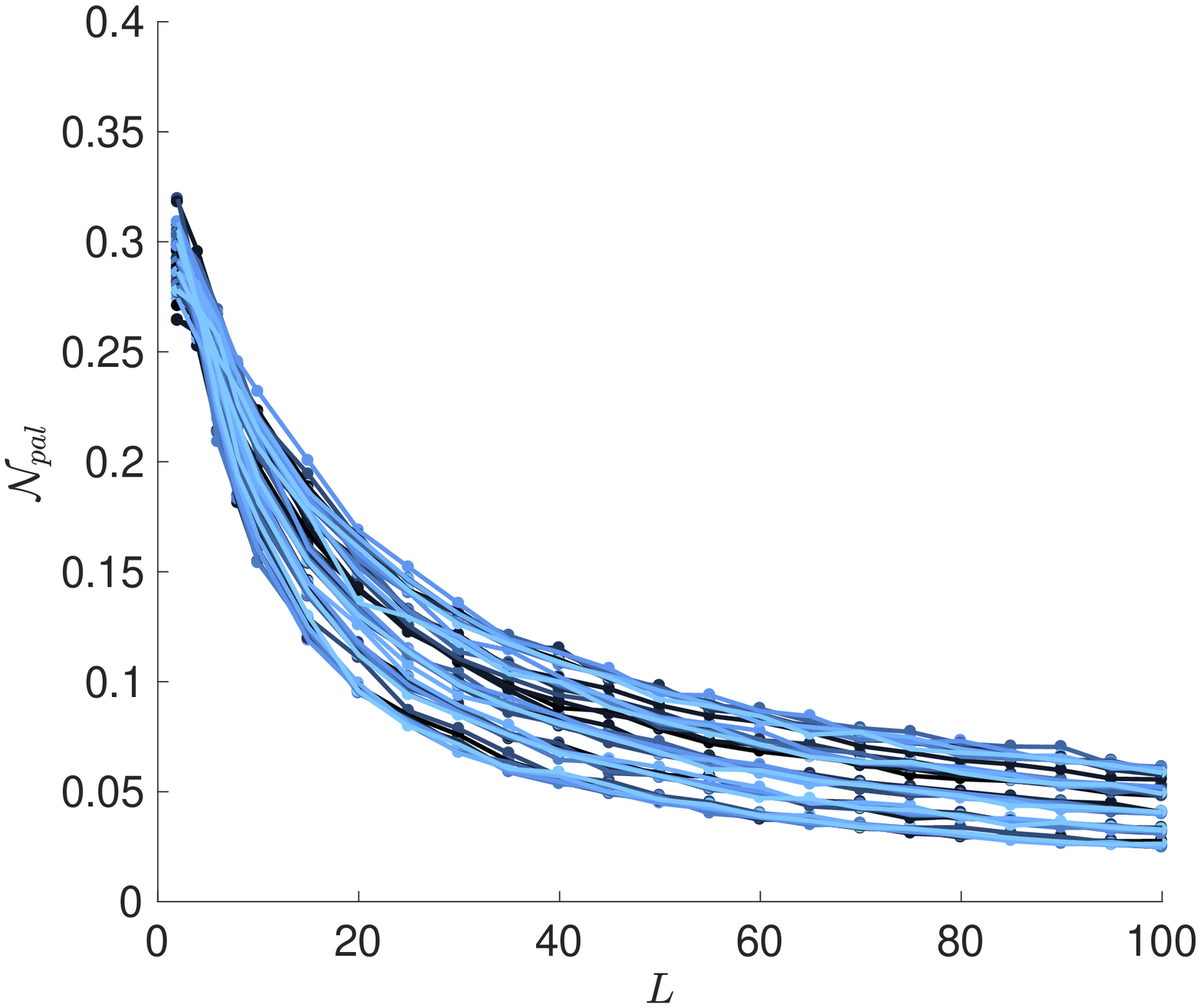}   \\ 
	        \multicolumn{4}{l}{{\bf (D)} }\\		
                \includegraphics[width=0.23\textwidth]{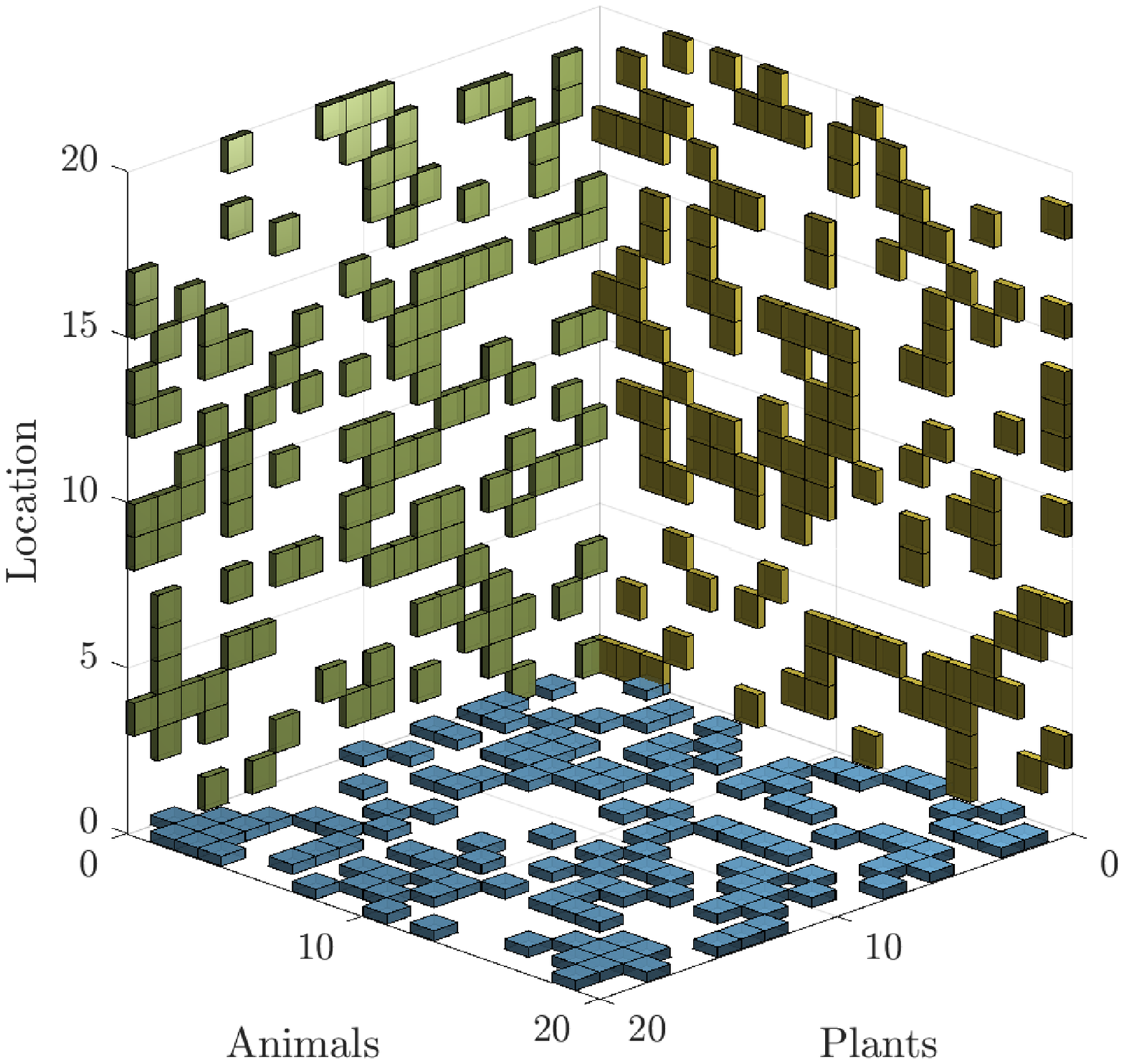} 
                & \includegraphics[width=0.23\textwidth]{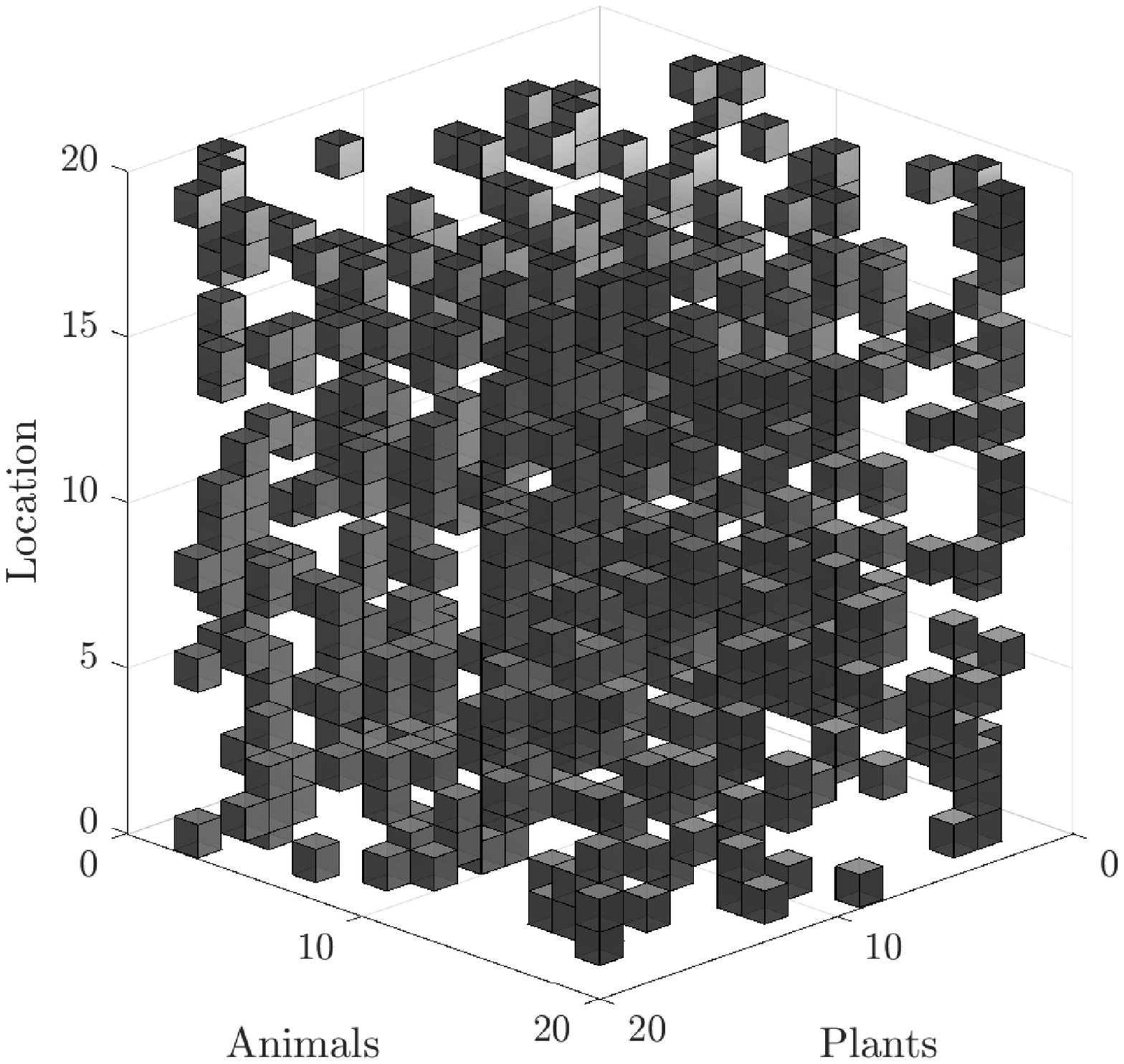} 
                & \includegraphics[width=0.23\textwidth]{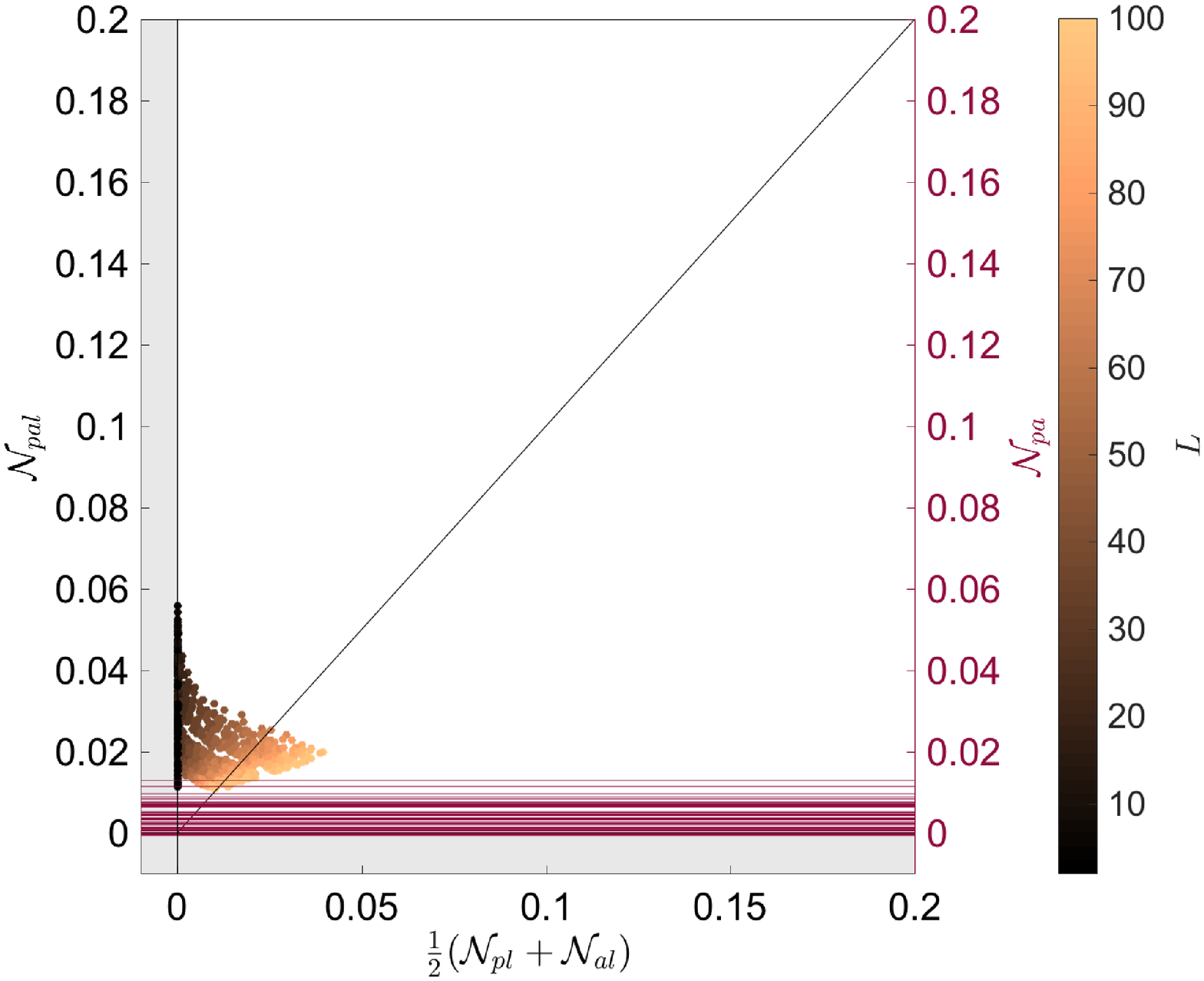}    
                & \includegraphics[width=0.21\textwidth]{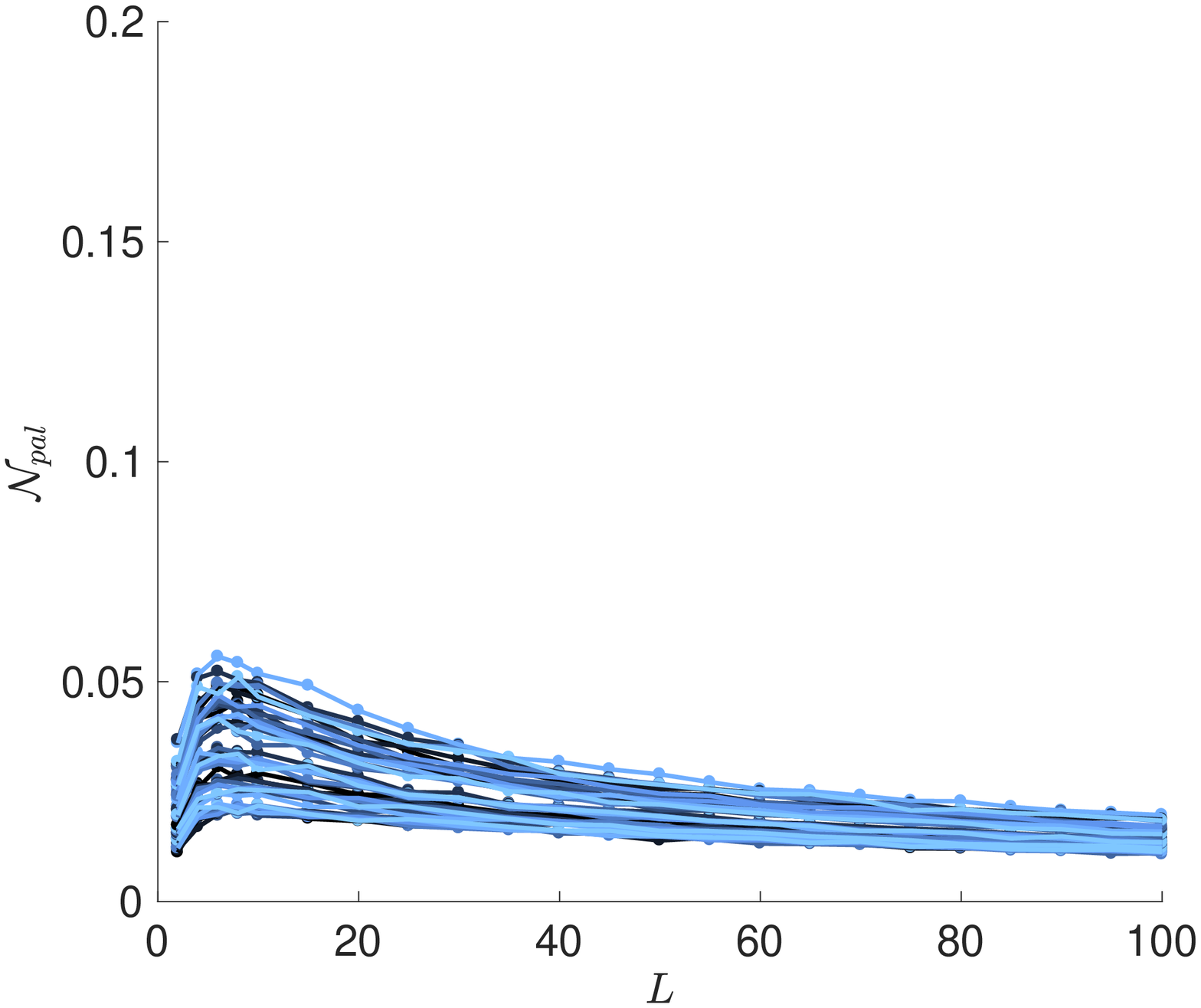}    

        \end{tabular}
        \caption{Four prototypical examples of $n$-dimensional hypercubes that describe tripartite relationships of an ecological system. This figure illustrates four limiting cases: those in which species-site occurrences are highly structured (A and B), against those in which such occurrences are arranged at random (C and D). The first column displays each of the three partial views of the system (plant-animal, $\mathcal{G}_{\subs{PA}}$; plant-location, $\mathcal{G}_{\subs{PL}}$; and animal-location, $\mathcal{G}_{\subs{AL}}$). The second column places in grey the presence of tripartite interactions. The aggregation of these hyperlinks on the projections defined by the axis reduces again to the view in the left column. The scatter plots in the third column inform about the relationship between the newly introduced $n$-dimensional nestedness (with $n=3$ here), $\mathcal{N}_{\subs{PAL}}$, and the ``traditional'' bipartite version of it. Among the observed patterns, we highlight that high levels of $\mathcal{N}_{\subs{PAL}}$ may appear despite low values of species-site overlap, i.e. a segregated scenario (panel C). Finally, the behaviour of $\mathcal{N}_{\subs{PAL}}$ against the granularity of the spatial dimension ($L$) is studied in the right-most column. Color code is proportional to $\xi_{\subs{PL}}$ and $\xi_{\subs{AL}}$. The brighter the color the larger $\xi$'s are. Black indicates $\xi$'s take the lower value for both projections. Lightest blue indicates $\xi$'s take the largest value for both projections. Noteworthy, we observe, for some arrangements, a non-monotonic evolution of $\mathcal{N}_{\subs{PAL}}$ --although by definition $\mathcal{N}_{\subs{PAL}}\sim 0$ for very large $L$. Results of third and fourth column correspond to networks created with parameters $\xi_{\subs{PA}} = 1$, $\alpha \in \{1, 1.5, 2, 2.5, 3\}$ and $c=5000$. See methods for a detailed explanation of the parameters.}
        \label{fig:toy}
\end{figure*}

Consider, without loss of generality on the dimensionality, a system composed of three dimensions of plants ($P$), animals ($A$), and locations ($L$), and their ternary relationships. That is, a tripartite hypergraph $\mathcal{G}_{\subs{PAL}}$, whose vertices are decomposed into three disjoint sets such that no two graph vertices within the same set are adjacent. $\mathcal{G}_{\subs{PAL}}$ encodes the species-species interactions (animal-plant projection), and the species-site occurrences (animal-location and plant-location projections), see left column in Figure~\ref{fig:toy}. Note that the hypergraph representation of ecological systems provides a natural way to extract pairwise relations between the ecological organisms, simply contracting the dimension we wish to omit, e.g. $a^{\subs{PA}}_{pa} = \sum_{\lambda=1}^L a^{\subs{PAL}}_{pa\lambda}$ where $a^{\subs{PA}}_{pa}$ and $a^{\subs{PAL}}_{pal}$ are the entries of the adjacency matrices representing $\mathcal{G}_{\subs{PA}}$ and $\mathcal{G}_{\subs{PAL}}$.

A high-dimensional representation demands a generalisation of the measures that are used to analyse bidimensional ecological systems\cite{pilosof2017multilayer}. Turning to nestedness, we propose here a natural extension of NODF\cite{almeida2008consistent}: the measure of connectivity overlap is augmented from rows and columns, to planes (for $n=3$) and hyperplanes (for $n > 3$). Noticeably, our measure  $\mathcal{N}_{\subs{PAL}}$ (and its contracted bipartite counterparts: $\mathcal{N}_{\subs{PA}}$, $\mathcal{N}_{\subs{PL}}$, $\mathcal{N}_{\subs{AL}}$) incorporate as well a null model term\cite{sole2018revealing} --see Methods for a complete formal description of the extensions we propose.

\paragraph{Idealised limiting scenarios.}  \label{par:synthetic} We first analyse how the structural patterns of the partial views of an ecological system relate to the structural patterns of the $n$-partite hypergraph organisation (note again that we restrict the analysis to $n=3$). Also, we provide here some intuitions on how nestedness evolves as a function of the third dimension of the system (locations in our example). To this end, we first develop a network generation model that, although oversimplified, already contains the main ingredients that will be analysed in the rest of the paper. Briefly, the model controls for the shape, the total connectance and the distribution of interactions on the final tripartite hypergraph, see Methods.


Figure~\ref{fig:toy} represents some synthetic extreme scenarios. In two of them (A and B), animal-plant interactions present a perfectly nested architecture ($\pi_{\subs{PA}} = 0$, see Methods), whereas both species-location matrices are either perfectly nested ($\pi_{\subs{AL}} = \pi_{\subs{PL}} = 0$; panel A), or totally random (panel B: $\pi_{\subs{AL}} = \pi_{\subs{PL}} = 1$, species can be found at any location with equal probability). In the other two (C and D), animal-plant interactions happen randomly ($\pi_{\subs{PA}} = 1$) while, again, animal-plant interactions are perfectly nested (C) or perfectly random (D). Note that, although all these configurations are plausible \cite{}, we do not claim any empirical resemblance of these settings to real communities, and we just use them as idealisations from which clear structural signatures arise.

It is important to highlight that, given any imposed pairwise relations, we can construct a hyper-dimensional network in any of the presented scenarios (Figure~\ref{fig:toy}, central column); i.e. the set of hyperedges is not empty. This stands as a (rather trivial) visual proof that competitive exclusion mechanisms (habitat shift\cite{connell80}, spatio-temporal niche segregation\cite{kronfeld2003partitioning}, etc.), which would lead to segregated species-site settings, are compatible with large species (plant-plant or animal-animal) niche overlap, i.e. nestedness (for example, panel C). Non-empty hypergraphs is the rule --rather than the exception-- against varying levels of randomness, density, and shape.

The third column of Figure~\ref{fig:toy} shows scatter plots for each extreme scenario (A-D). In those plots, $N_{\subs{PAL}}$ in the y-axis is compared to the average of species-site nestedness ($(N_{\subs{AL}} + N_{\subs{PL}})/2$) in the x-axis. As a visual aid, the ``classical'' $N_{\subs{PA}}$ is marked as horizontal red lines as well. Each point in the scatter plot corresponds to a single network. Such networks have been generated for a fixed set of parameters (i.e. limiting scenarios), with varying $L$. Quite notably, only in one scenario (B) $N_{\subs{PAL}} < (N_{\subs{AL}} + N_{\subs{PL}})/2$ for almost all cases, regardless of the number of sites of the tripartite graph. Such setting corresponds to a disordered species-species interaction, combined with a strictly nested species-site pattern. Other than that, A and C show that, for $L \lesssim 40$, $n$-dimensional systems, $N_{\subs{PAL}} > (N_{\subs{AL}} + N_{\subs{PL}})/2$ --this is particularly true for C, in which species display nested interaction together with site segregation. In other words, this setting corresponds to an ecosystem in which exploiting similar resources is common (nestedness in the projected $\mathcal{G}_{\subs{PA}}$ graph), but at the same time competitive exclusion mechanisms are in place (segregation in the $\mathcal{G}_{\subs{PL}}$ and $\mathcal{G}_{\subs{AL}}$ projections). Panel D stands as a baseline setting --all relationships are random--, with negligible values for system-wide and partial measures of nestedness.

We now turn to the sensitivity of $n$-dimensional nestedness $\mathcal{N}_{\subs{PAL}}$ to the granularity of the spatial division (Figure~\ref{fig:toy}, right column). In ecological terms, maintaining a fixed ecological area of study and increasing the granularity of the {\it location} dimension, controls which spatial scale most affects the interaction among species. Larger ${L}$ implies a finer geographical resolution, which results in a smaller probability of any two given species to be observed in the same site. This would be equivalent to go from a system where spatial distances that affect interaction among species is large\cite{owen2015spatially} (small $L$) to a system where this distances are small \cite{albrecht2001spatial} (large $L$).

Intuitively, one might expect that increasing $L$ renders, in turn, smaller $\mathcal{N}_{\subs{PAL}}$. Indeed, it can be shown that $\mathcal{N}_{\subs{PAL}} \to 0$ as ${L} \to \infty$. However, experiments show that $\mathcal{N}_{\subs{PAL}}$ is not always monotonic with respect to $L$. The only observed situation where increasing $L$ decreases $\mathcal{N}_{\subs{PAL}}$ for any $\Delta L$ is the scenario where the $\mathcal{G}_{\subs{PA}}$ is nested, whereas $\mathcal{G}_{\subs{PL}}$ and $\mathcal{G}_{\subs{AL}}$ are unstructured (panel C). In all other cases we observe an increase in the value of $\mathcal{N}_{\subs{PAL}}$ before starting a monotonic decrease towards $0$. Recall that $\mathcal{N}_{\subs{PAL}}$ reduces to  $\mathcal{N}_{\subs{PA}}$ when $L = 1$. The precise $L$ at which such maximum is reached varies with the particular network features (particularly: size and density; see Supplementary Information), but are otherwise similar. In this sense, this simple example suffices to show that $\mathcal{N}_{\subs{PAL}}$ may have well-defined bounds, and these are related to the nestedness and segregation levels observed at the aggregated 2-dimensional projections.



This concave shape might have important implications in the ecological systems. On one hand, it evidences that only having information about the relationship between plant-animal is not sufficient to understand the overall structure of the system, since $\mathcal{N}_{\subs{PA}}$ can underestimate or overestimate $\mathcal{N}_{\subs{PAL}}$ depending on the spatial scale of the system. That is to say, $\mathcal{N}_{\subs{PA}}$ is a bad predictor of the system-wide overlap level. On the other hand, observing a maximum on $\mathcal{N}_{\subs{PAL}}$ implies that there may be an optimal spatial scale, probably related to species' characteristic range\cite{sexton2009}, which should contribute to the dynamics running on the system. 

These results (third and fourth column of Figure~\ref{fig:toy}) are robust against different network connectance and weight distribution fixed on the $\mathcal{G}_{\subs{PA}}$ projection. See the Supplementary Information where results for $5,000$ (Section S1) and $50,000$ (Section S2) links, and different interaction distribution strategies, are evaluated.




\begin{figure*}[h]
    	\begin{tabular}{ll}
                \includegraphics[width=0.5\textwidth]{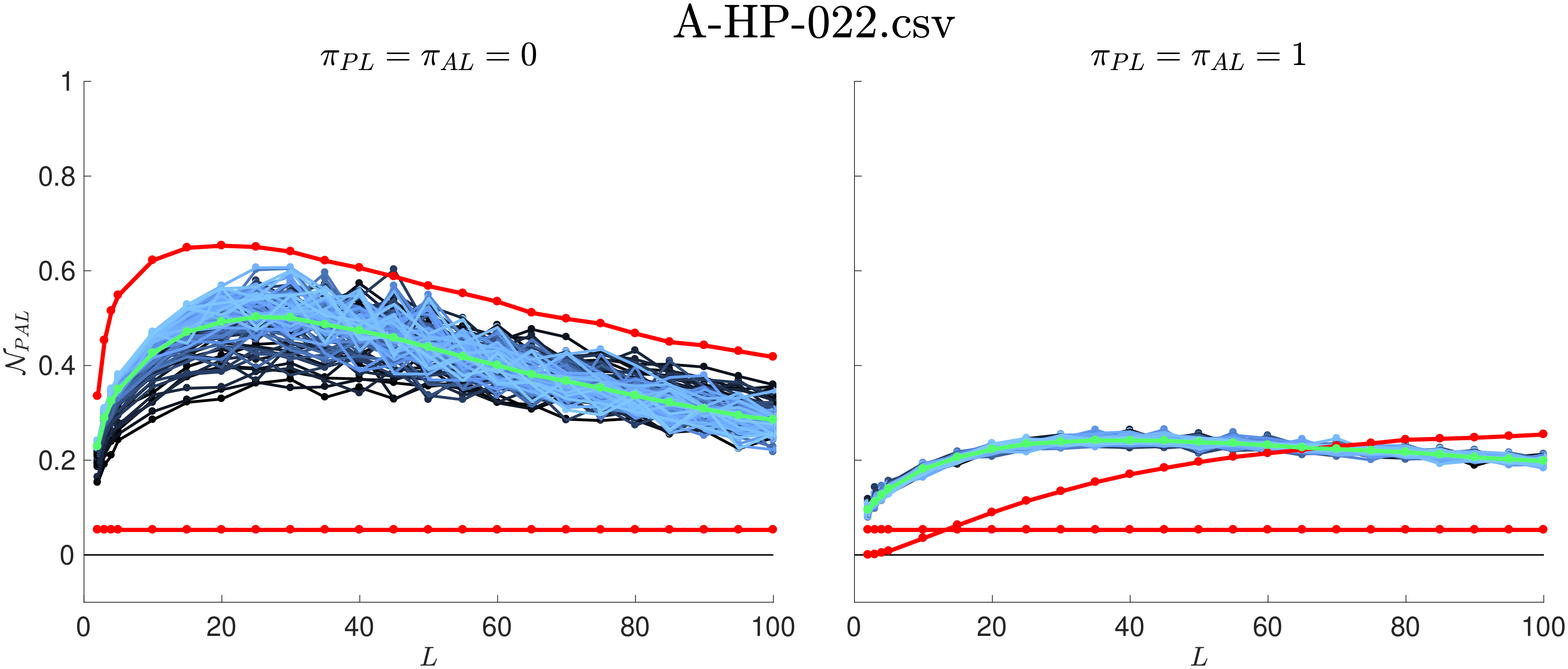} 
                & \includegraphics[width=0.5\textwidth]{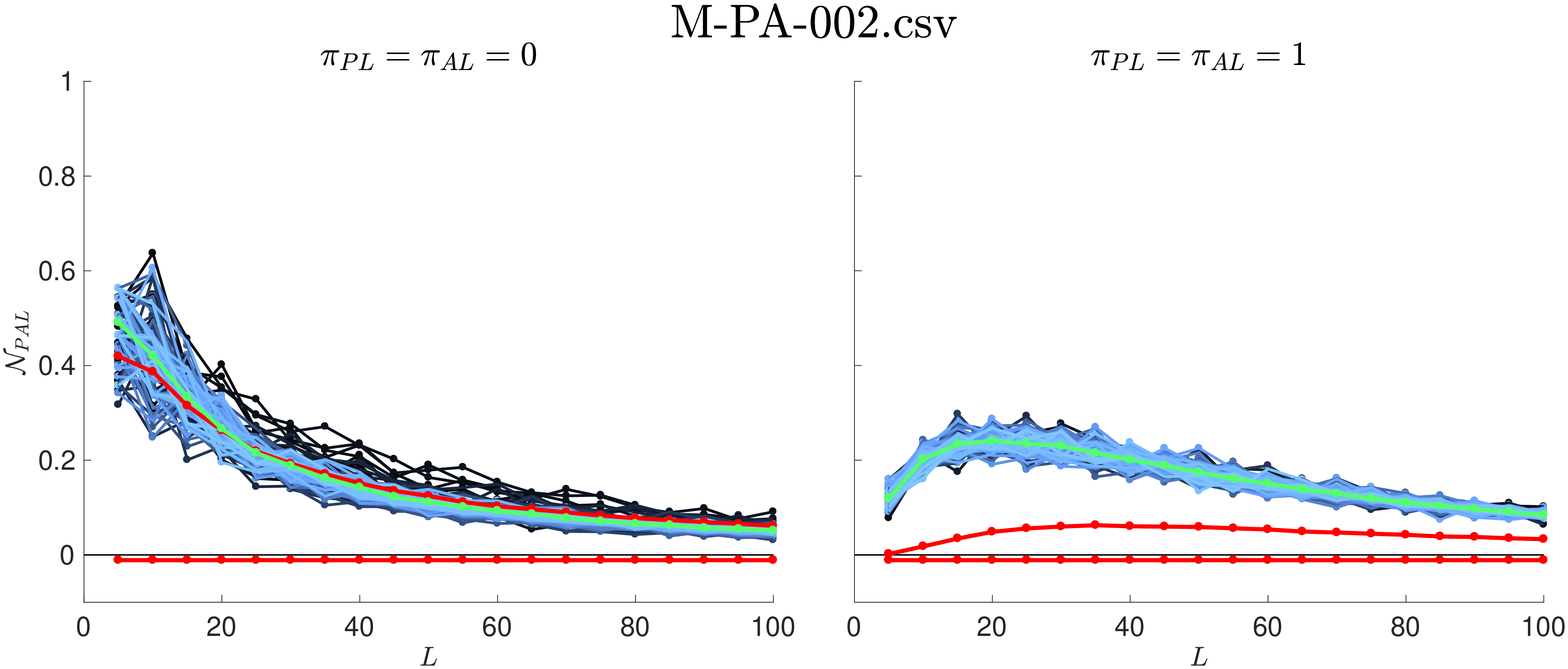} \\
		\includegraphics[width=0.5\textwidth]{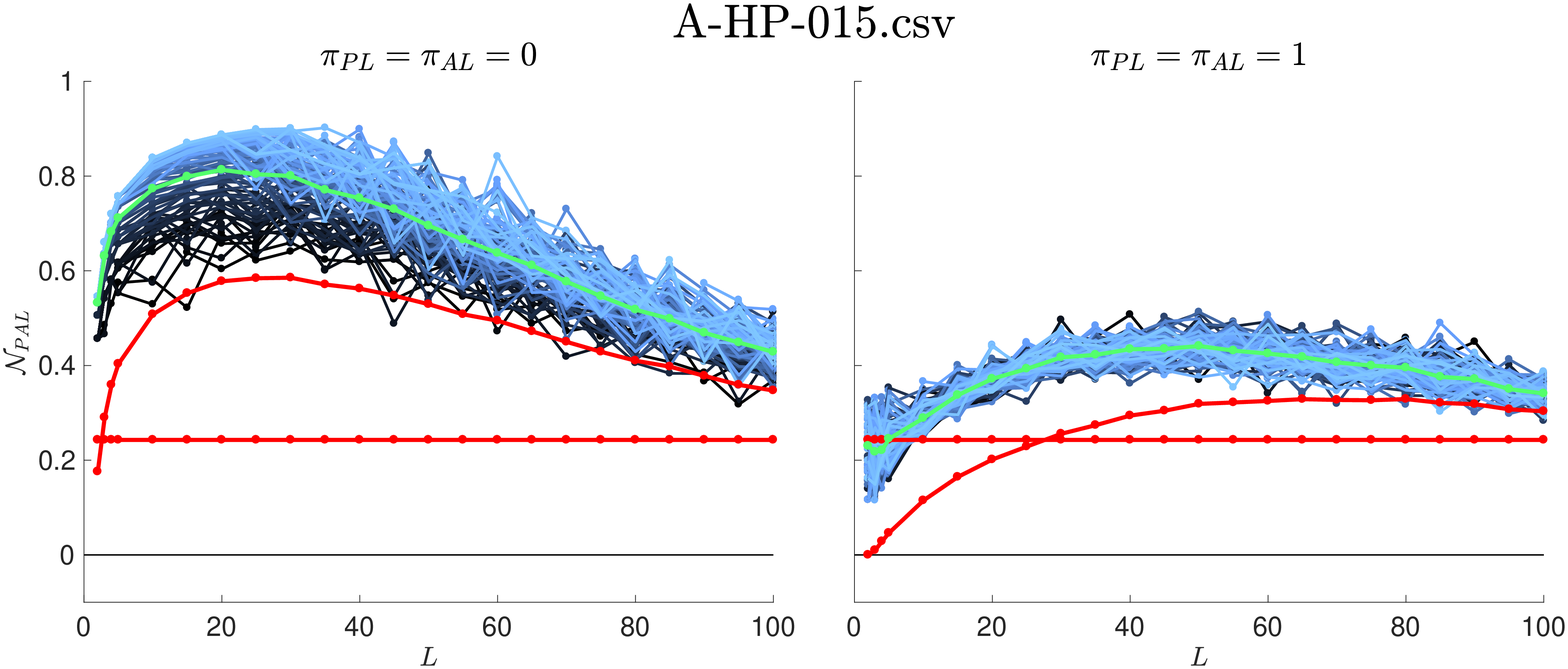} 
		& \includegraphics[width=0.5\textwidth]{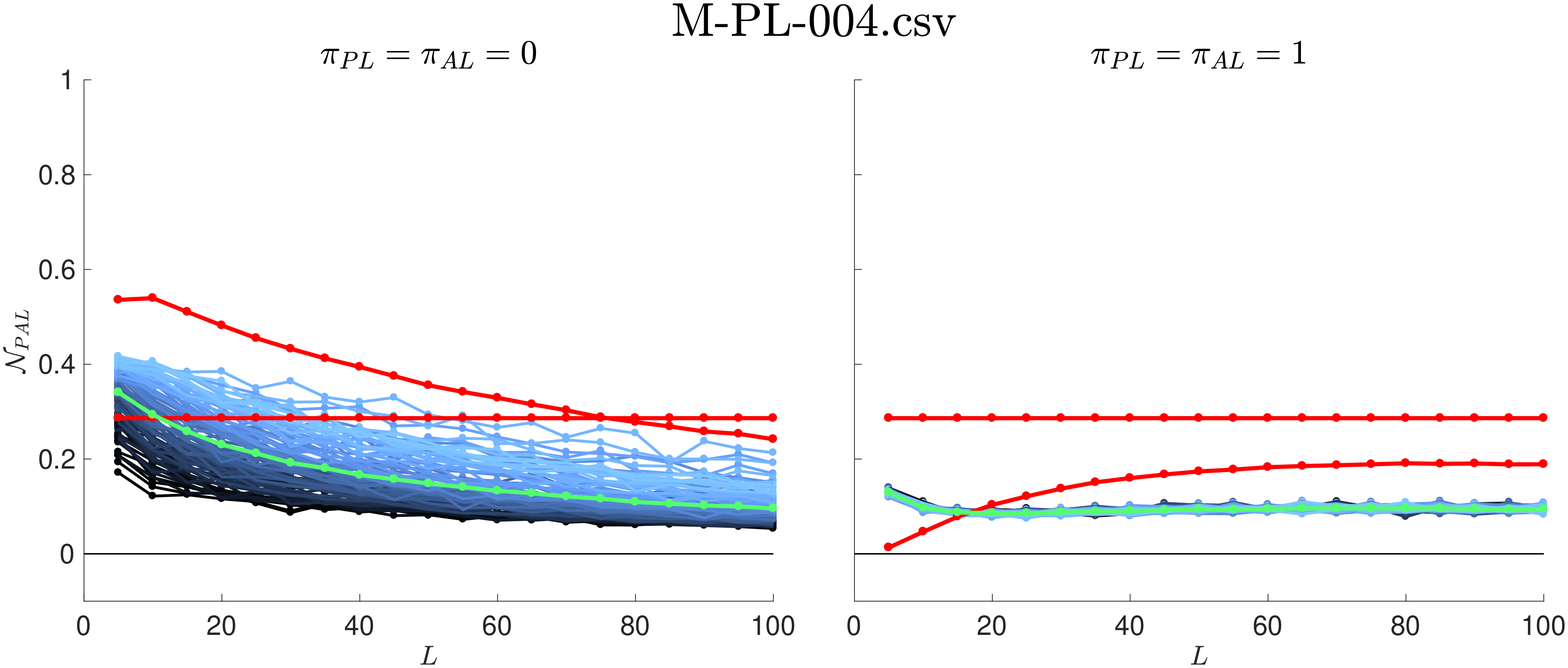} \\
                \includegraphics[width=0.5\textwidth]{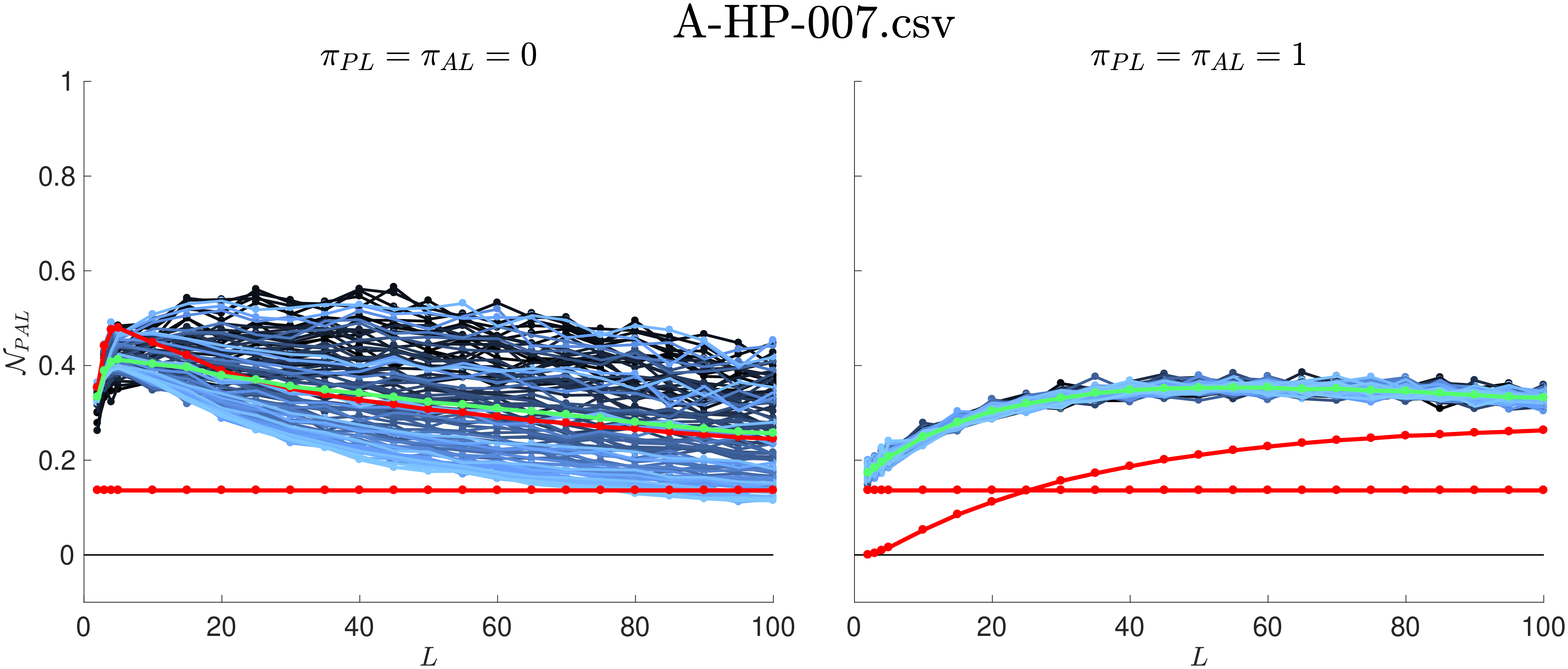}  
                & \includegraphics[width=0.5\textwidth]{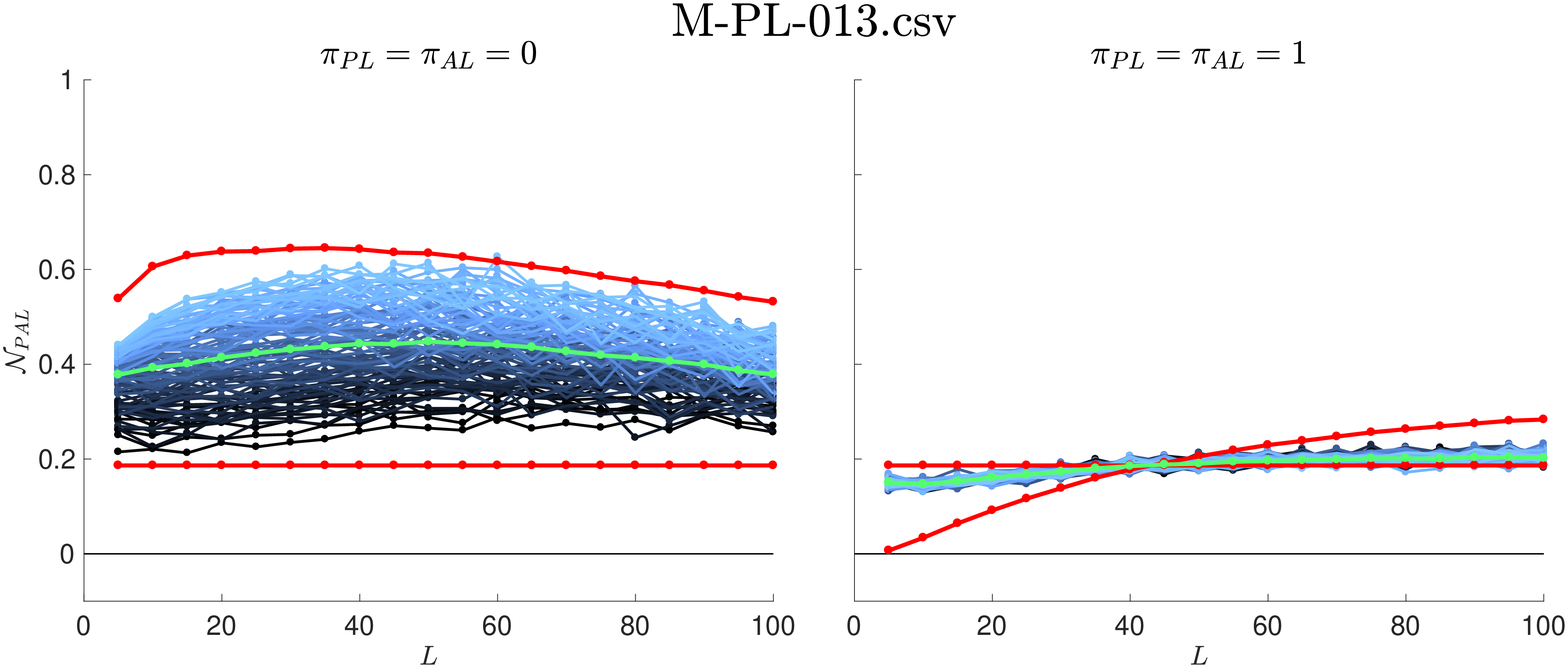} \\
        \end{tabular}
        \caption{Behaviour of $\mathcal{N}_{\subs{PAL}}$ against $L$ in six hybrid (empirical-synthetic) hypergraphs. For each dataset (3 host-parasite and 3 pollination networks)\cite{weboflife}, we have generated a tripartite hypergraph that incorporates the spatial dimension in two extreme cases: first and third columns correspond to settings in which the spatial dimension ($\mathcal{G}_{\subs{PL}}$ and $\mathcal{G}_{\subs{AL}}$) have been arranged as perfectly nested, $\pi_{\subs{PL}} = \pi_{\subs{AL}} = 0$; in the second and fourth columns, those projections present a random disposition, $\pi_{\subs{PL}} = \pi_{\subs{AL}} = 1$. Comparing these two settings for the same dataset, we observe that $\mathcal{N}_{\subs{PAL}}$ is, in most cases, larger than $\mathcal{N}_{\subs{PA}}$ (solid red line, which corresponds to a measurement on real data, i.e. the $\mathcal{G}_{\subs{PA}}$), no matter how the spatial dimension is laid out. And yet, this is not a systematic observation, which remarks the idea that a partial knowledge of the system (nestedness in the species-species interactions) is not a reliable proxy to the system-wide overlap.}
        \label{fig:empirical}
\end{figure*}

\paragraph{Potential scenarios in synthetic-empirical networks.} \label{par:real} We now exercise the previous scheme on a large set of empirical, bipartite weighted networks which correspond to mutualistic pollination and host-parasite communities\cite{weboflife}. Location data is unavailable and consequently species-site relationships are missing. However, the number of interactions and interaction distributions among species (which we continue to name $\mathcal{G}_{\subs{PA}}$) are fixed by the observed data. Methods section shows how we construct hyper-graphs compatible with these data. Realistic parameters of the projections ($\mathcal{G}_{\subs{PL}}$ and $\mathcal{G}_{\subs{AL}}$ projections) are totally unknown (they depend on the ecological system we consider), and so we explore all the parameter space. 

Figure~\ref{fig:empirical} plots the evolution of $\mathcal{N}_{\subs{PAL}}$ as a function of ${L}$  for four real datasets. For each dataset (41 in total), $4\times 10^{3}$ species-site networks are generated either in a fully ordered ($\pi_{\subs{PL}} = \pi_{\subs{AL}} = 0$) or fully random ($\pi_{\subs{PL}} = \pi_{\subs{AL}} = 1$) manner. Different density levels have been tested as well (note that we know from data the density in the $\mathcal{G}_{\subs{PA}}$, but not for the rest of projections). The dashed red line corresponds to $\mathcal{N}_{\subs{PA}}$, which is obtained from empirical data and remains constant, as it is independent of $L$. The solid red line corresponds to $(N_{\subs{AL}} + N_{\subs{PL}})/2$. Black lines correspond to the evolution of $\mathcal{N}_{\subs{PAL}}$ from individual network realisations, while in cyan we observe the aggregate 3-dimensional nestedness for the whole ensemble. What stems from this Figure is that the analysis of the $\mathcal{G}_{\subs{PA}}$ provides limited information about the structure of the $n$-dimensional ecological system. From these selected examples, it is clear that the results are comparable to the ones in Figure~\ref{fig:toy}: the analysis of $\mathcal{N}$ on a single projection (and, specifically, the $\mathcal{G}_{\subs{PA}}$) can under- or over-estimate  $\mathcal{N}_{\subs{PAL}}$, failing in most cases as a reliable proxy of $n$-dimensional nestedness of the system. 

\begin{figure*}[h]
    	\begin{tabular}{ll}
                (A) & (B) \\	
                \includegraphics[width=0.5\textwidth]{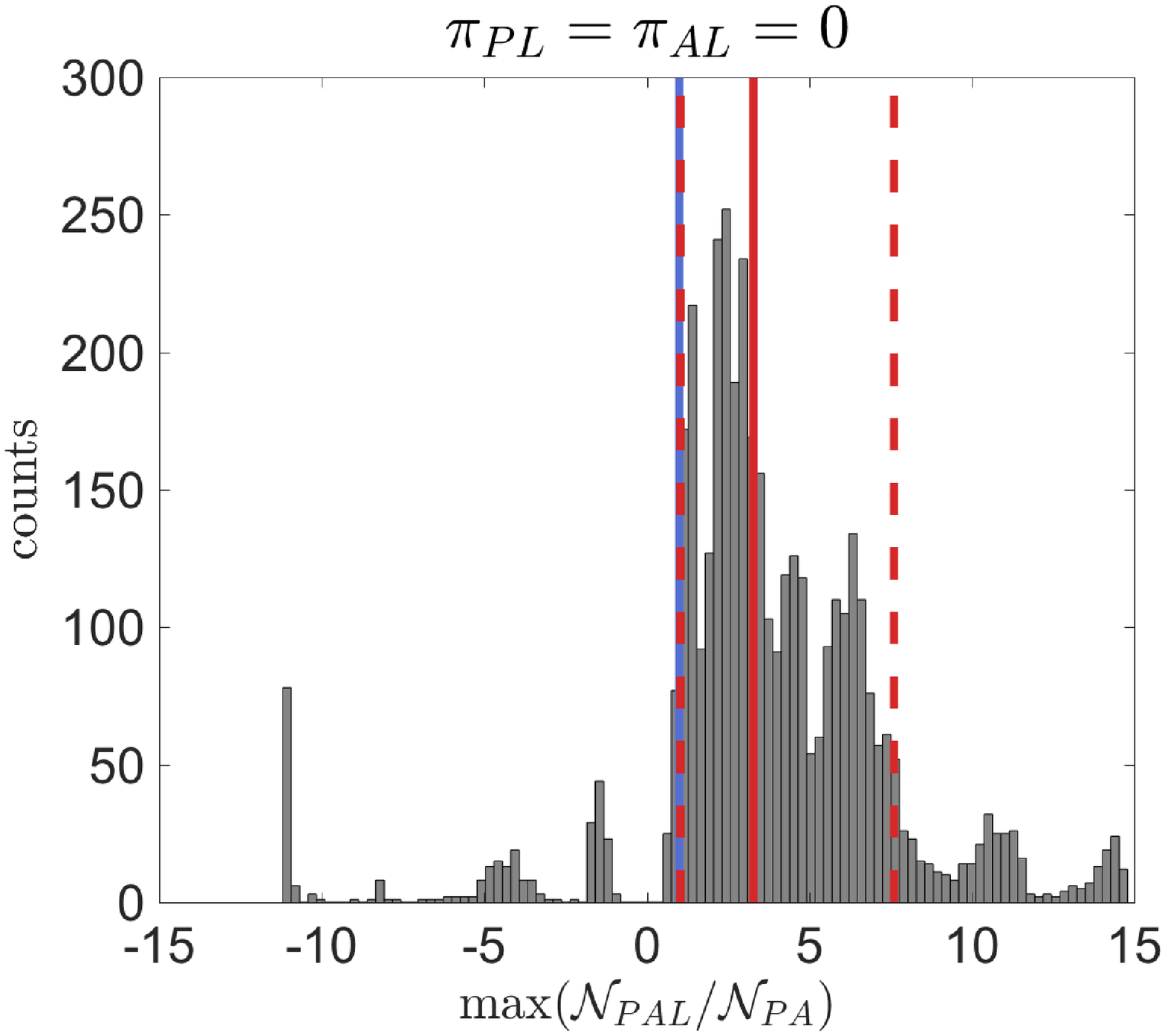} 
                & \includegraphics[width=0.5\textwidth]{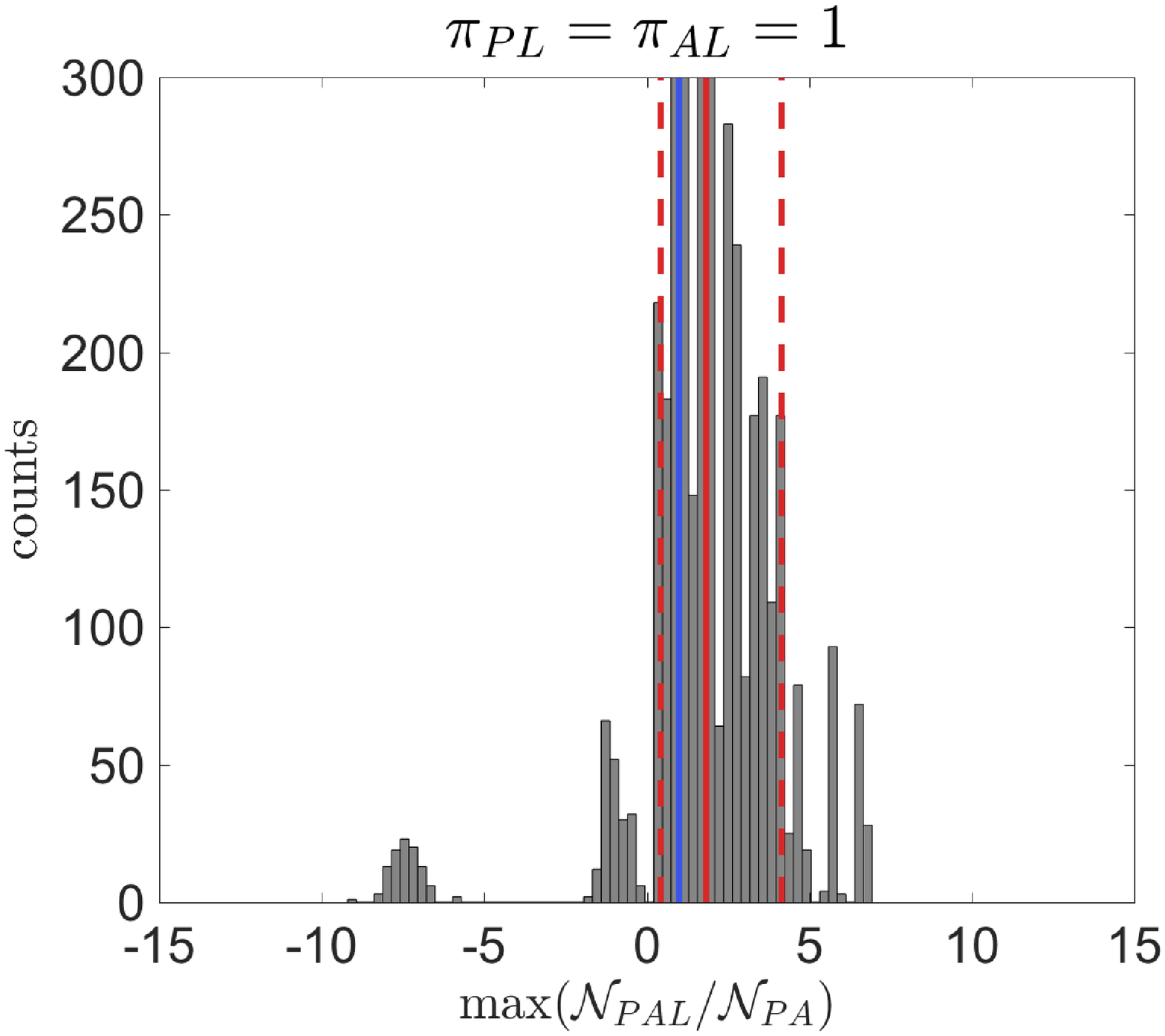} \\
        \end{tabular}
        \caption{Histogram with the ratio between 3-dimensional nestedness ($\mathcal{N}_{\subs{PAL}}$) and the bipartite nestedness $\mathcal{N}_{\subs{PA}}$ as measured from real plant-animal or host-parasite data. Each count in the histograms corresponds to the maximum ratio between those quantities, which in turn depends on the number of locations ($L$). Panel (A) shows the results for experiments where species-site relationships are generated as perfectly nested ($\pi_{\subs{PL}} = \pi_{\subs{AL}} = 0$); in panel (B), on the contrary, $\pi_{\subs{PL}} = \pi_{\subs{AL}} = 1$. Red dashed vertical lines correspond to first and ninth decile, and the red solid line marks the median. The blue vertical line marks, as a visual aid, $\max (\mathcal{N}_{\subs{PAL}} / \mathcal{N}_{\subs{PA}}) = 1$. Negative values appear rarely, when one of the nestedness values of the ratio is negative, i.e. when nestedness is less than the expected at random.}
        \label{fig:histogram}
\end{figure*}

This last conclusion is also visible in Figure~\ref{fig:histogram}. Panel A and B report a summary of the maximum ratio between the obtained $\mathcal{N}_{\subs{PAL}}$ and $\mathcal{N}_{\subs{PA}}$ among the generated hybrid (half empirical, half synthetic) tripartite hypergraphs. 
The amount of times when $\mathcal{N}_{\subs{PAL}}=\mathcal{N}_{\subs{PA}}$, i.e. their ratio equals 1 (highlighted in blue), is small. Also, leaving aside extreme results (beyond the first and ninth decile, dashed red lines), the system-wide 3-dimensional nestedness is most often larger than its bipartite counterpart (which is an empirical measurement, provided that the $\mathcal{G}_{\subs{PA}}$ corresponds to actual data). This is true regardless of the dominant pattern in $\mathcal{G}_{\subs{PL}}$ and $\mathcal{G}_{\subs{AL}}$. This has important consequences: although system-wide data is not available (in which species-species and species-site informations comes together), we foresee, through this extensive exploration of potential scenarios, that system-wide nestedness will be higher than observed 2-dimensional nestedness. Negative values in the histogram indicate that one of the nestedness measures is lower than the expected one, i.e. the null model term is larger than the observed overlap.

\begin{figure*}[h]
    	\begin{tabular}{ll}
                (A) & (B) \\		
                \includegraphics[width=0.5\textwidth]{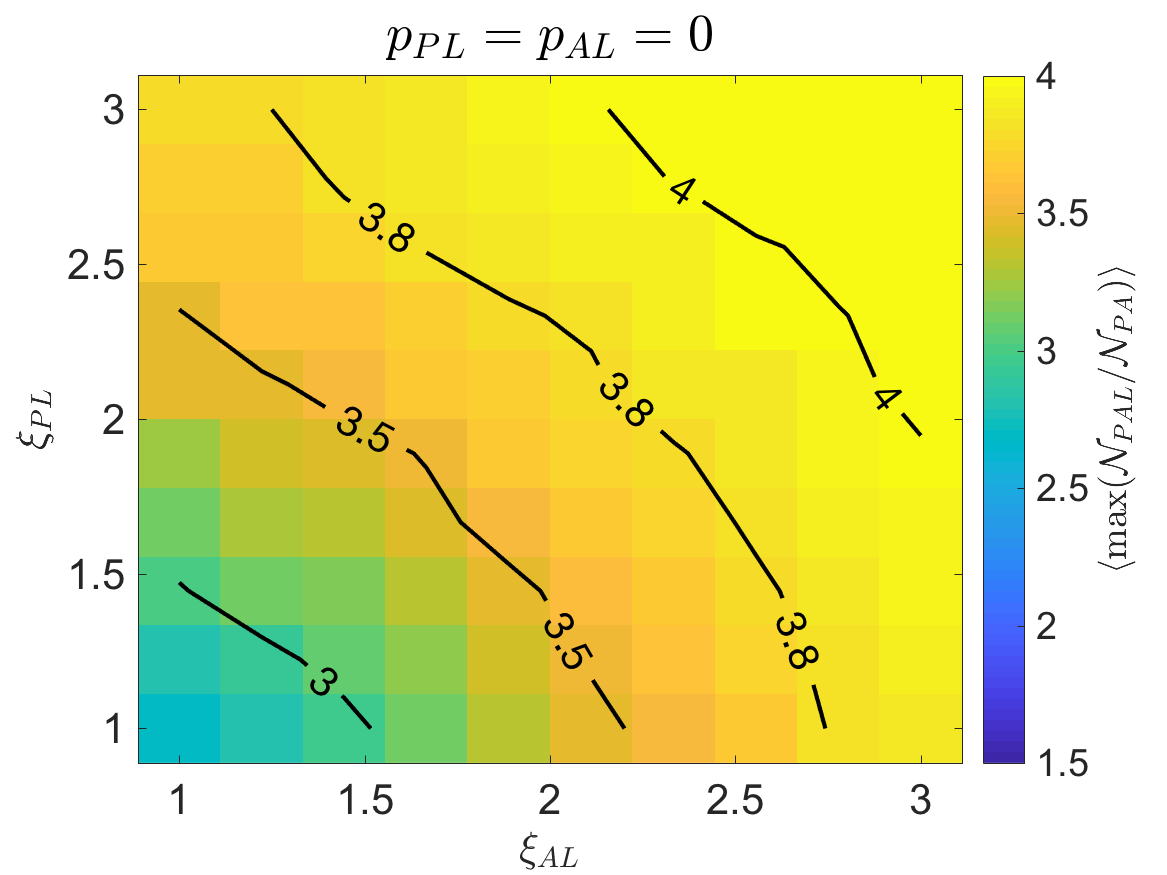} 
                & \includegraphics[width=0.5\textwidth]{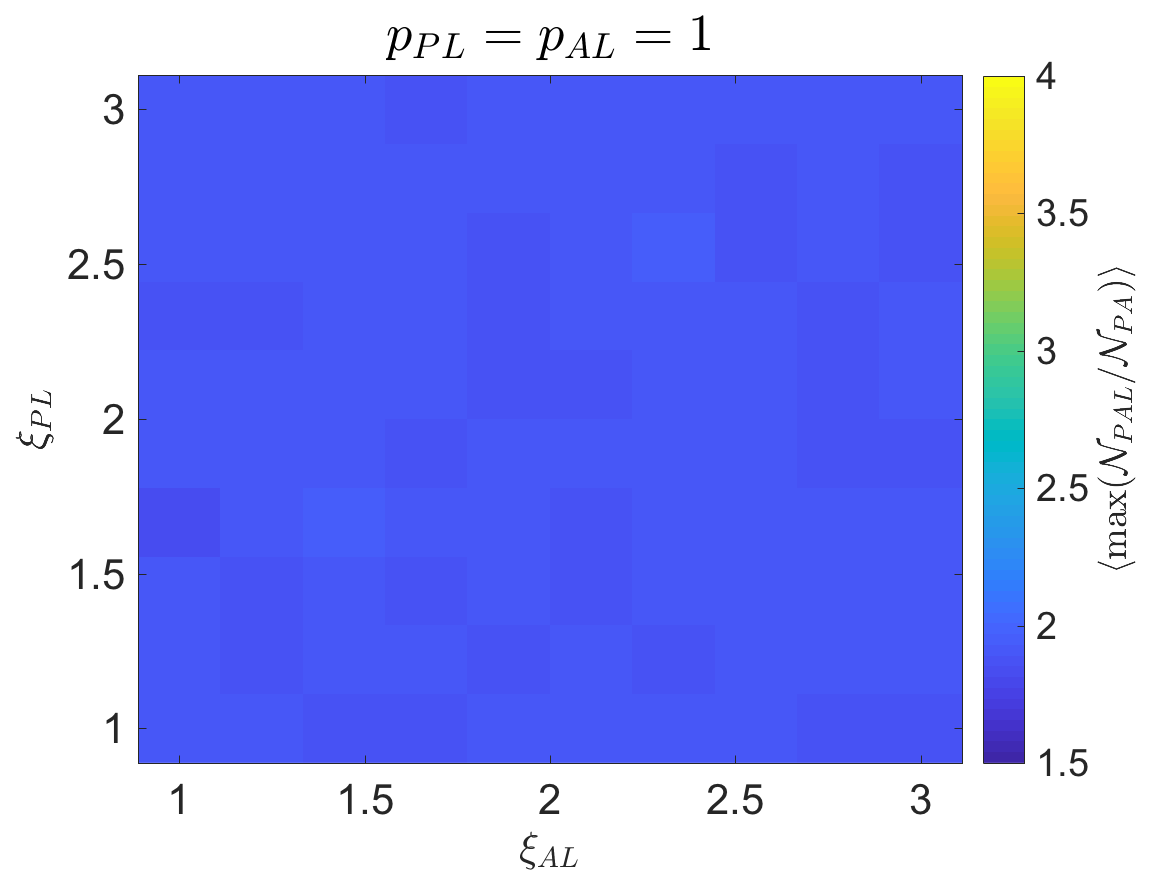} \\	
                (C) & (D) \\	
                \includegraphics[width=0.5\textwidth]{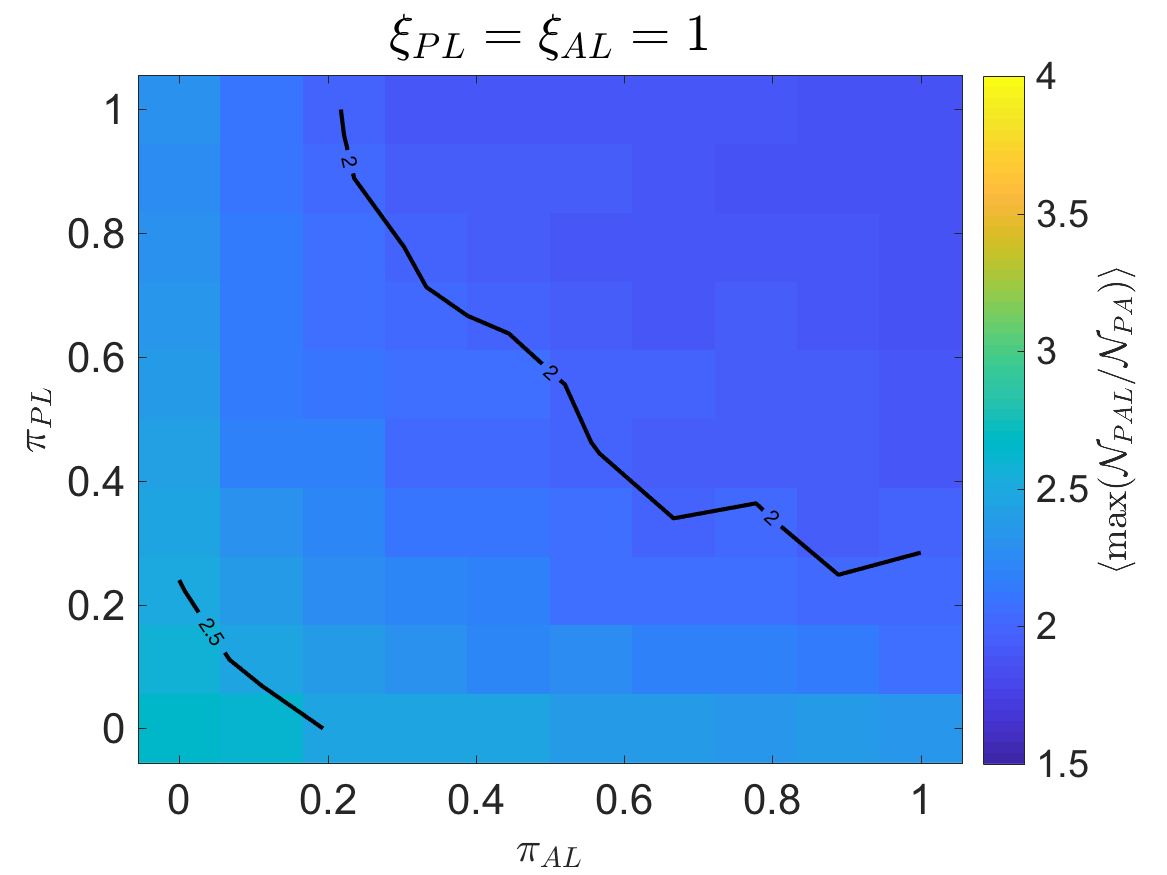} 
                & \includegraphics[width=0.5\textwidth]{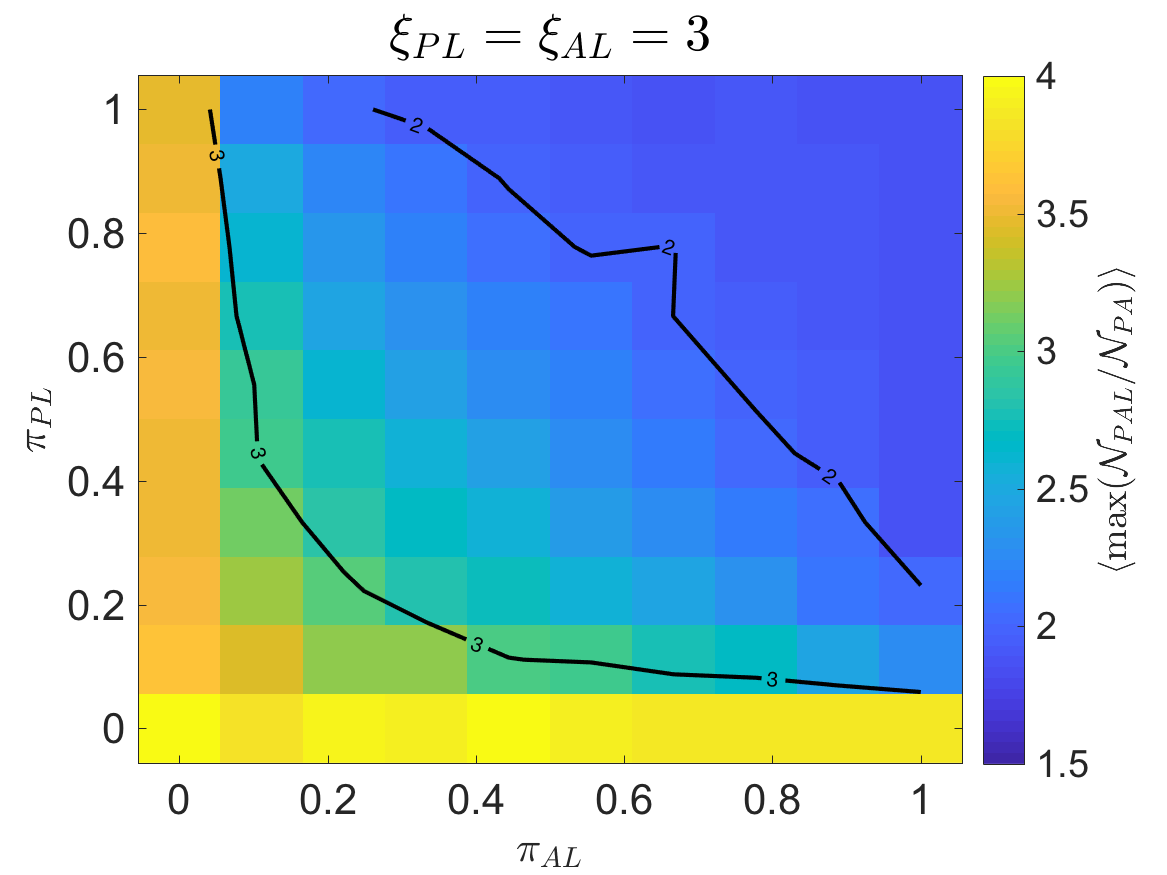} \\
        \end{tabular}
        \caption{Diagram with the relationship between 3- and 2-dimensional nestedness (the second one being empirically measurable from real datasets) with respect to the network generation parameters, $\xi$ and $p$, see Methods. Each value in the diagrams corresponds to the maximum ratio, fixing noise $p$ and shape $\xi$ of the synthetic layers (species-location), varying the length of the location dimension $L$. Panels (A) and (B) show the results with respect to varying values of shape for both projections and, panel (C) and (D) with respect to noise. The colormap (common to all panels) already indicates that $\max (\mathcal{N}_{\subs{PAL}} / \mathcal{N}_{\subs{PA}}) > 1$, whatever combination of parameters we choose. This is particularly remarkable for panel (B), in which the species-location projections present a random pattern.}
        \label{fig:heatmaps}
\end{figure*}

Eventually, Fig. \ref{fig:heatmaps} shows the influence of the system parameters on the accuracy of $\mathcal{N}_{\subs{PA}}$ to estimate $\mathcal{N}_{\subs{PAL}}$, in terms of the maximum ratio between the obtained $\mathcal{N}_{\subs{PAL}}$ and $\mathcal{N}_{\subs{PA}}$. Panels A and B show the results with respect to the shape ($\xi_{\subs{AL}}$ and $\xi_{\subs{PL}}$) of the pair-wise projections (see Methods). The accuracy is in general low for non-noisy systems and becomes minimum when the shape is low for both projections. As noise ($\pi$) increases, the accuracy improves, with the ratio (at the limit) achieving $1.8$ for all settings (panel B). Recall that $\pi$ controls the noise, and so $\xi$ merely controls the density of the network for large $\pi$ values. Panels C and D show the same results but in terms of $\pi$. We, again, see that accuracy improves when both projections lack of internal structure ($\pi \approx 1$).


\section{Discussion}
The need to harmonise apparently diverging drivers in an ecosystem's dynamics --nestedness and segregation-- has triggered so far different strategies. Adaptive foraging offers a very reasonable approach: in a generally nested setting, the best strategy for a pollinator is to exploit its most exclusive resource. That is, the most generalist pollinator is pushed to visit the most specialist plants, to avoid in this way competition with those competitors ranked below it. This is a logical outcome to escape the contradicting principles under discussion, if one remains within the constraints of a 2-dimensional view of an ecosystem. Here we have presented the formalism to upscale the representation of a niche, inspired by the original vision of Hutchinson's $n$-dimensional space. Such formalism comes under the form of a $n$-partite hypergraph for which, following the recent literature on multilayer networks, a useful set of descriptors and methods can be defined --descriptors and methods which parsimoniously generalise those already known. 
Along this line, we introduce the concept of $n$-dimensional nestedness, which is the way to quantify the overall amount of overlap in a system, and which does not reduce to any trivial combination of ``traditional'' (bipartite) nestedness of the partial projections of such system. 
At the face of a data gap, we have deployed such measure in fully and half synthetic examples, to show that seemingly irreconcilable mechanisms may coexist in an ample range of parameters. Additionally, we have shown that the analysis of a partial view of the system provides, in general, a poor approximation to overall features. To some extent, this result is expected in an upscaled system that implies an increment in the degrees of freedom --but conveys, as well, higher complexity.

This work, which contains many simplifying assumptions, would then be the onset for many questions that can be now reframed. Future work, within the vision of an $n$-dimensional representation of ecosystems, should tackle the lack a well-grounded theory to determine how large $n$ should be for that representation to be most explicative, beyond suggestive heuristics\cite{eklof13} in similar questions. This problem is adjacent to the challenges around the collection of the necessary data, which admittedly can be resource-intensive. Data gathering should, ideally, grow in breadth (including, beyond species-species interactions, at least spatial and temporal information); and in depth, moving from the dominant binary (presence/absence) paradigm towards the consideration of weights, which may have a significant impact on the topological measures\cite{staniczenko2013ghost}. The existence of optimal spatiotemporal scales --as suggested by our results in simplified scenarios-- adds further complexity to these challenges.

From a purely structural perspective several possibilities arise. Closely related to this work, one could consider multiple scenarios for the proposed synthetic model, which have been disregarded here. Note that our example networks (e.g. Fig.\ref{fig:toy}A) make a strong assumption when correlating the generalist/specialist conditions to all the projections. That is, our synthetic hypergraphs exclude, for example, the possibility that a plant can be generalist (animal-wise) and specialist (location-wise) at the same time, which is admittedly an arbitrary limitation. More generally, a high-dimensional view of ecosystems reignites the debate around which are the dominant topological patterns in such systems\cite{fortuna2010nestedness,thebault2010stability,valverde2018architecture}, which demands in turn an methodological upscaling --think for example in the wide variety of patterns reviewed in\cite{lewinsohn2006structure,presley2010comprehensive}, which have been laid out in a two-dimensional framework. While some of such developments are in place (e.g. centrality measures\cite{sole2014centrality}, modularity and community detection in multilayer settings\cite{pilosof2017multilayer}, etc.), others are largely missing.
The structural perspective leads naturally to the functional one: beyond the question over which architecture(s) is (are) prevalent, the ultimate challenge is to understand why. This points directly to the dynamical aspects of the problem, related to the growth and assembly mechanisms that have driven the system to a given high-dimensional structure; and to the persistence of the system, and its connection to certain topological arrangements that promote stability above other configurations.



\section{Methods}
\label{sec:methods}

\subsection{Nestedness on $\mathbf{n}$-dimensional sets}
\label{sec:3dnes}

Nestedness is a classically structural measure used to study the relationship between two, usually disjoint and independent, sets of ecological organisms $P$ and $A$  (e.g. $P~:=~ \mbox{`plants'}$ and $A~:=~\mbox{`animals'}$). Given the relational nature of the data, a bipartite network, $\mathcal{G}_{\subs{\subs{PA}}}~=~\{ P \cup A,\mathcal{E}_{\subs{\subs{PA}}}\}$, is the convenient mathematical object to consider disregarding, in turn, within-organism relations. This simplification restricts the possible edges on the bipartite structure to $\mathcal{E}_{\subs{\subs{PA}}} \in \{P \times A\}$. This has been the usual approach with some exceptions \cite{pilosof2017multilayer,gracia2017joint}.

Definitely, if the dimensionality of the ecological system is larger than two \cite{eklof13}, the bipartite approach provides only a partial view -- a projection from a higher dimensional space as we will show -- of the full ecological system. State-of-the-art tools for nestedness analysis do not allow for the computation of nestedness in larger dimensional systems. To overcome such dimensionality restrictions, we first require a equivalent higher dimensional structure such as $n$-partite hypergraphs where the set of all essential dimensions of the system can be represented: 
\begin{equation}
	\mathcal{G}_{\subs{D_1 D_2 \dots D_n}}= \{D_1 \cup D_2 \cup \dots \cup D_n, \mathcal{E}_{\subs{D_1 D_2 \dots D_n}} \}
\end{equation}
where, equivalently to the bi-dimensional case, relations within the two sets may be disregarded and only edges (or hyper-edges) between the sets can exist $\mathcal{E}_{\subs{D_1 D_2 \dots D_n}}~\in~\{D_1~\times~D_2~\times~\dots~D_n \}$. A prototypical example of a system with $n=3$ could be $\mathcal{G}_{\subs{\subs{PAL}}}$, where $P:=\mbox{`plants'}$, $A:=\mbox{`animals'}$ and $L:=\mbox{`locations'}$. The hypergraph object provides a natural way to extract pairwise relations between the ecological dimensions. Consider the previous example $\mathcal{G}_{\subs{\subs{PAL}}}$, as in \cite{owen2015spatially}. We can extract pair-wise interactions by contracting the dimension we wish to ignore or, what is the same compute the marginals against this dimension: 
\begin{equation}
	\mathcal{G}_{\subs{\subs{PA}}} = \sum_{\ell=1}^{\card{L}} \mathcal{G}_{\subs{\subs{PA\ell}}} \label{eq:projection}
\end{equation}
A convenient way to represent the network $\mathcal{G}_{\subs{\subs{PAL}}}$ is in terms of an $n$-dimensional hypercube (the equivalent of an adjacency matrix in a higher dimensional space) $\mathcal{A}$, where the elements $\mathcal{A}_{p_1a_1l_1}$ indicate the strength of the relation between plant $p_1$ and animal $a_1$ at location $l_1$. As in the presence-absence matrix $\mathcal{A}_{p_1a_1l_1} = 1$ might indicate that the relation exist and $\mathcal{A}_{p_1a_1l_1}=0$ that the relation does not exist. 

With this new object at hand, we are now in position to extend the nestedness analysis to $n$-dimensional ecological systems. There are many measures for nestedness analysis \cite{ulrich2009consumer}, for convenience we will focus on the NODF measure \cite{almeida2008consistent} over the unweighted structure (i.e. presence-absence matrices). As in the 2-dimensional case, for the computation of the $n$-dimensional nestedness we will proceed by fixing each of the dimensions of the system and then compute the local contribution with respect to the remaining dimensions. Thus, in a network with $n$ dimensions $\mathcal{G}_{\subs{D_1D_2\dots D_n}}$, $\mathcal{N^{\subs{D_1}}}$ will indicate the contribution to the total nestedness of dimension $D_1$ with respect to $D_2D_3\dots D_n$.  Eventually the total NODF can be obtained by averaging the local contribution for each set,
\begin{equation}
	\widetilde{\mathcal{N}} = \frac{\sum_{i=1}^{n} \card{D_i}\mathcal{N}^{\subs{D_i}}}{\sum_{i=1}^{n} \card{D_i} } 
	\label{eq:global3DNODF}
\end{equation}
Noteworthy, contributions for each dimension in eq. \ref{eq:global3DNODF} are weighted linearly with respect to the cardinality of the set they represent. This differs from the original NODF \cite{almeida2008consistent}, where the contribution of each dimension are weighted quadratically with respect to the cardinality of the set. Quadratic weightings can be problematic on networks where the cardinalities of the different sets are large: in those settings, the nestedness value depends mostly on the largest sets, and this might not be convenient. Instead, a linear weighting offers a more natural approach.

The computation of the local contribution $\mathcal{N}^{\subs{D_i}}$ requires extending NODF's first and most basic ingredient, the {\it node overlap}. Consider first a 3-dimensional system, $\mathcal{G}_{\subs{\subs{PAL}}}$, fixing two nodes $p_1$ and $p_2$ from the same set $P$, the overlap $\mathcal{O}^P_{p_1p_2}$ is related to the amount of commonly shared neighbours on the remaining dimensions of the system. Formally, this can be obtained as 
\begin{equation}
	\mathcal{O}^{\subs{P}}_{p_1p_2} = \sum\limits^{\card{A}}_{i}\sum\limits^{\card{L}}_{j} \mathcal{A}_{p_1ij} \mathcal{A}_{p_2ij}
\end{equation}
And, from here, given two nodes $u_1$ and $u_2$ of set $D_u$, in a $n$-dimensional system
\begin{equation}
	\mathcal{O}^{\subs{D_u}}_{u_1 u_2} = \sum\limits^{\card{D_1}}_{j_1=1}\dots\sum\limits^{\card{D_i}}_{\substack{j_i=1\\ i \ne u}} \dots \sum\limits^{\card{D_n}}_{j_n=1} \mathcal{A}_{j_1\dots j_i \dots u_1 \dots j_n} \mathcal{A}_{j_1\dots j_i \dots u_2 \dots j_n}
\end{equation}
Besides the overlap the rest of ingredients of $\widetilde{\mathcal{N}}^{D_i}$ correspond mainly to normalisation factors 
\begin{equation}
	\widetilde{\mathcal{N}}^{\subs{D_i}} = \frac{2}{\card{D_i}(\card{D_i}-1)}\sum\limits^{|D_i|}_{u,v = 1} \frac{\mathcal{O}^{\subs{D_i}}_{uv}}{k_v} \Theta(k^{\subs{D_i}}_u - k^{\subs{D_i}}_v)\,
	\label{eq:local3DNODF}
\end{equation}
where the Heaviside function controls that the degree $k^{\subs{D_i}}_u$ of node $u$ is larger than the degree of node $v$ as in the original formulation of NODF. Equivalently to the 2-dimensional case, $k^{\subs{D_i}}_u$ corresponds to the number of links connected to node $u$ on dimension $D_i$. That is, 
\begin{equation}
	k^{\subs{D_i}}_u = \sum\limits^{\card{D_1}}_{j_1=1}\dots\sum\limits^{\card{D_i}}_{\substack{j_i=1\\ i \ne u}} \dots \sum\limits^{\card{D_n}}_{j_n=1} \mathcal{A}_{j_1\dots j_i \dots u \dots j_n}
\end{equation}	
As in the bipartite case, one possible concern is the consideration of nodes with equal degree on the computation of the NODF. In eq. \ref{eq:local3DNODF} this translates into what value we assign to Heaviside function at zero.

A $n$-partite hypergraph will be perfectly nested if, for each set $D_i$, every pair of nodes ($u$,$v$) with different degree ($k^{D_i}_u > k^{D_i}_v$), the edges connecting $v$ are a subset of the edges connecting $u$. The degree of the involved nodes influences their overlap: the larger their degree, the higher their probability to exhibit overlap by chance. To balance this undesired effect, it is convenient to introduce a null model that compensates for the intrinsic chance of overlap. Along the lines of \cite{sole2018revealing}, we propose to subtract, from the observed overlap, the expected overlap that nodes might exhibit by chance. For convenience considering first a 3-partite hypergraph over $P$, $A$ and $L$ sets, given two nodes $p_{1}$ and $p_{2}$ of set $P$, their expected overlap is
\begin{equation}
\langle O^{P}_{p_1p_2} \rangle = \sum\limits^{|A|}_{i}\sum\limits^{|L|}_{j} \frac{k^{\subs{P}}_{p_1} k^{\subs{P}}_{p_2}}{{\left(\card{A} \card{L}\right)}^2} = \frac{k^{\subs{P}}_u k^{\subs{P}}_v}{\card{A} \card{L}}
\end{equation}
We refer the reader to \cite{sole2018revealing} for a detailed explanation of the expected overlap in a 2-dimensional case.

From here it is not difficult to obtain the expected overlap for the $n$-dimensional setting. Given two nodes $u_1$ and $u_2$ of set $D_u$ it corresponds to
\begin{align}
\langle O^{D_u}_{u_1u_2} \rangle = &\sum\limits^{\card{D_1}}_{j_1=1}\dots\sum\limits^{\card{D_i}}_{\substack{j_i=1\\ i \ne u}} \dots \sum\limits^{\card{D_n}}_{j_n=1} \frac{k^{\subs{D_u}}_u k^{\subs{D_u}}_v}{{\left[\prod\limits_{j=1}^n\card{D_j}\right]}^{n-1}} \\
	= &\frac{k^{\subs{D_u}}_u k^{\subs{D_u}}_v}{{\prod\limits_{j=1}^n\card{D_j}}}
\end{align}

Eventually the extension of NODF to $n$-partite hypergraphs with dimensions $D_1$ to $D_n$, including a null model, becomes
\begin{align}
\mathcal{N}=&\\ =&\frac{2}{\sum\nolimits_{i=1}^{n}\card{D_i}}  \sum\limits_{i=1}^{n}\sum\limits^{|D_i|}_{u,v = 1} \frac{\mathcal{O}^{\subs{D_i}}_{uv} - \langle\mathcal{O}^{\subs{D_i}}_{uv}\rangle}{k^{\subs{D_i}}_v(|D_i|-1)} \Theta(k^{\subs{D_i}}_u - k^{\subs{D_i}}_v) \nonumber
\end{align}
%
%
%
%
%
%
%
%
%
%
%
%

\subsection{Generation mechanisms of n-partite hypergraphs}

The section describes the mechanism we propose to generate $n$-partite hypergrahs that describe the interaction of organisms in an ecological system. Since the experiments are performed considering only 3 dimensions, we stick to that dimensionality for the ease of formulation and understanding . However, the extension to $n$-partite hypergraphs is straightforward. In the following, graph $\mathcal{G}_{\subs{PAL}}$ (and $a^{\subs{PAL}}_{pal}$ its adjacency matrix elements $(p,a,l)$) will stand for hypothetical interaction data between plants and animals at a location. Similarly, $\mathcal{G}_{\subs{PA}}$ (and $a^{\subs{PA}_{pa}}$ its adjacency matrix elements $(pa)$) stands for the pair-wise marginal interaction between the same organisms.

The datasets in section \ref{sec:res} are generated considering location probabilities of plants, $P^L_p(\ell)$, and animals, $P^L_a(\ell)$, with respect to $L$ dimension and number of observed interactions between plants and animals, $\mathcal{G}_{\subs{PA}}$. The only difference between the generation of synthetic (section \ref{sec:res} a) and real (section \ref{sec:res} b) datasets lies on how the bipartite relation $\mathcal{G}_{\subs{PA}}$ is obtained: for real systems, it is experimentally given by \cite{weboflife}; whereas for synthetic systems it is artificially constructed, see \ref{sec:pa_proptodegree}.

For the construction of the datasets, we are first interested in obtaining the probability that two species interact in a given location $\ell$, $P^{I}_{pa}(\ell)$, to then sample proportionally the observed interactions based on the observations $\mathcal{G}_{\subs{PA}}$. To do so, we define $P^{I}_{pa}(\ell)$ to be proportional to the probability that the two species are found in the same location. Formally, $P^{I}_{pa}(\ell) = \kappa P^{L}_{pa}(\ell)$, where $\kappa$ is a normalisation constant and is obtained as $\kappa = 1/\sum_{\lambda=1}^{\card{L}} P^{L}_{pa}(\lambda)$. Without further knowledge about the influence of animal locations given the location of plants, we assume their location is independent of each other. Thus, $P^{L}_{pa}(\ell)=P^{L}_{p}(\ell)P^{L}_{a}(\ell)$ and so $P^{I}_{pa}(\ell) = \kappa P^{L}_{p}(\ell)P^{L}_{a}(\ell)$.

Then, considering $\mathcal{A}_{\subs{PA}}$ as the average expected number of interactions between species independently we have that $\langle a^{\subs{PAL}}_{pal} \rangle = a_{pa}P^{I}_{pa}(\ell)$. However, to allow some variation on the distribution of the interactions, instead of rounding $\langle a^{\subs{PAL}}_{pal} \rangle$, we opted to generate network samples by randomly distributing the amount of observed interactions ($a_{pa}$) among the different $\ell$ considering a multinomial distribution with probabilities in $P^{I}_{pa}(\ell)$.

\subsection{Generation of synthetic $\mathcal{G}_{\subs{PA}}$} \label{sec:pa_proptodegree}

The main ingredients to generate an instance of $\mathcal{G}_{\subs{PA}}$ are $\xi$, $\pi$, and total network connectance $c$. The first parameter $\xi$ controls for the shape (``slimness'') of a purposefully nested structure. On the other hand, $\pi$ mimics random and uncorrelated noise, removing links from the perfectly nested structure. That is, parameter value $\pi = 0$ corresponds to the initial structure (perfect nestedness, shape $\xi$), and $\pi = 1$ corresponds to an Erd{\H{o}}s-R\'{e}nyi network. See section III A of \cite{sole2018revealing} for a detailed description. 
We first choose which organism pairs can interact considering interaction probabilities $P^{I}_{pa}$. This resumes to throwing a random number $r$ and locate a link in $\hat{a}^{\subs{PA}}_{pa}$ if $P^{I}_{pa}<r$. Then, among the chosen links, we distribute total amount of interactions $c$ according to two strategies. In the first strategy we distribute $c$, as a function of the degree of the organisms. Thus, given that $k_p$ and $k_a$ are the degree of pants and animals obtained through $A_{\subs{PA}}$, the observed interactions between $p$ and $a$ are given by
\begin{equation}
	a^{\subs{PA}}_{pa} = c\frac{(k_p k_a)^\alpha}{\sum_{ij}(k_i k_j)^\alpha}
	\label{eq:prop_to_degree_strategy}
\end{equation}
where $\alpha>1$ controls the amount of interactions between large degree nodes and low degree nodes. For $\alpha=1$ the distribution of interactions will be proportional to de degree of organisms and for large $\alpha$ values, interactions will tend to be mainly between large degree organisms.

The second strategy is a particular case of \ref{eq:prop_to_degree_strategy} where interactions are uniformly distributed among all the organisms that can interact. This can be obtained by, taking $\alpha=0$ and defining $0^0 = 0$.

%

\end{document}